\documentclass[10pt]{aastex}
\usepackage{amsmath,natbib}
\usepackage{rotating}
\usepackage{emulateapj5}

\newcommand{\eV}{{\,\rm eV}}
\newcommand{\nrun}{n_{\rm run}}

\newcommand{\Mpc}{\text{Mpc}}
\newcommand{\Gyr}{\text{Gyr}}

\newcommand{\CAMB}{\textsc{camb}}
\newcommand{\COSMOMC}{\textsc{CosmoMC}}
\newcommand{\BOOMERANG}{B03}
\newcommand{\XFASTER}{\textsc{Xfaster}}
                   
\slugcomment{submitted to ApJ}

\shorttitle{BOOMERANG Parameters}
\shortauthors{MacTavish et al.}

\begin{document}

\title{Cosmological Parameters from the 2003 flight of BOOMERANG}

\author{C.~J.~MacTavish\altaffilmark{1}, P.~A.~R.~Ade\altaffilmark{2},
J.~J.~Bock\altaffilmark{3}, J.~R.~Bond\altaffilmark{4},
J.~Borrill\altaffilmark{5}, 
A.~Boscaleri\altaffilmark{6}, P.~Cabella\altaffilmark{7}, 
C.~R.~Contaldi\altaffilmark{4,}\altaffilmark{8},
B.~P.~Crill\altaffilmark{9}, P.~de~Bernardis\altaffilmark{10}, 
G.~De~Gasperis\altaffilmark{7}, A.~de~Oliveira-Costa\altaffilmark{11}, 
G.~De~Troia\altaffilmark{10}, G.~Di~Stefano\altaffilmark{12},
E.~Hivon\altaffilmark{9}, A.~H.~Jaffe\altaffilmark{8},
W.~C.~Jones\altaffilmark{13}, 
T.~S.~Kisner\altaffilmark{14},A.~E.~Lange\altaffilmark{13}, 
A.~M.~Lewis\altaffilmark{4}, S.~Masi\altaffilmark{10}, P.~D.~
Mauskopf\altaffilmark{2}, 
A.~Melchiorri\altaffilmark{10,}\altaffilmark{15}, T.~E.~Montroy\altaffilmark{14},
P.~Natoli\altaffilmark{7,}\altaffilmark{16},
C.~B.~Netterfield\altaffilmark{1,}\altaffilmark{17},
E.~Pascale\altaffilmark{1}, F.~Piacentini\altaffilmark{10}, 
D.~Pogosyan\altaffilmark{4,}\altaffilmark{18}, G.~Polenta\altaffilmark{10},
S.~Prunet\altaffilmark{19}, S.~Ricciardi\altaffilmark{10}, G.~Romeo\altaffilmark{12},
J.~E.~Ruhl\altaffilmark{14}, P.~Santini\altaffilmark{10},  
M.~Tegmark\altaffilmark{11}, M.~Veneziani\altaffilmark{10},
and N.~Vittorio\altaffilmark{7,}\altaffilmark{16} }

\altaffiltext{1}{Department of Physics, University of Toronto, ON, Canada}
\altaffiltext{2}{School of Physics and Astronomy, Cardiff University,
Wales, UK}
\altaffiltext{3}{Jet Propulsion Laboratory, Pasadena, CA, USA}
\altaffiltext{4}{Canadian Institute for Theoretical Astrophysics (CITA),
University of Toronto, ON, Canada}
\altaffiltext{5}{National Energy Research Scientific Computing Center,
LBNL, Berkeley, CA, USA}
\altaffiltext{6}{IFAC-CNR, Firenze, Italy}
\altaffiltext{7}{Dipartimento di Fisica, Universit\`a di Roma Tor Vergata,
Roma,
Italy }
\altaffiltext{8}{Department of Physics, Imperial College, London, UK}
\altaffiltext{9}{Infrared Processing and Analysis Center, California
Institute of Technology, Pasadena, CA, USA}
\altaffiltext{10}{Dipartimento di Fisica, Universit\`a di Roma La Sapienza,
Roma,
Italy}
\altaffiltext{11}{Department of Physics, Massachusetts Institute of
Technology, Cambridge, MA, USA}
\altaffiltext{12}{Istituto Nazionale di Geofisica e Vulcanologia, Roma,
Italy}
\altaffiltext{13}{Department of Physics, California Institute of
Technology, Pasadena, CA, USA}
\altaffiltext{14}{Department of Physics, Case Western Reserve
University, Cleveland, OH, USA}
\altaffiltext{15}{INFN, Sezione di Roma 1, Roma, Italy}
\altaffiltext{16}{INFN, Sezione di Roma 2, Roma, Italy}
\altaffiltext{17}{Department of Astronomy and Astrophysics, University of
Toronto, ON, Canada}

\altaffiltext{18}{Department of Physics, University of Alberta, Edmonton,
AB, Canada}
\altaffiltext{19}{Institute d'Astrophysique de Paris, Paris, France}

\begin{abstract}
We present the cosmological parameters from the CMB intensity and
polarization power spectra of the 2003 Antarctic flight of the BOOMERANG
telescope.  The BOOMERANG data alone constrains the parameters of the
$\Lambda$CDM model remarkably well and is consistent with
constraints from a multi-experiment combined CMB data set.  We add LSS data from the
2dF and SDSS redshift surveys to the
combined CMB data set and test several extensions to the standard model
including:  running of the spectral index, curvature, tensor modes, the
effect of massive neutrinos, and an effective
equation of state for dark energy.  We also include an analysis of
constraints to a model which allows a CDM isocurvature admixture.

\end{abstract}


\section{Introduction}

The angular power spectra of the Cosmic Microwave Background (CMB) have become
invaluable observables for constraining cosmological models.  The
position and amplitude of the peaks and dips of the CMB spectra
are sensitive to such parameters as the geometry of
the Universe, the cosmological constant and the energy densities associated
with baryons and cold dark matter \citep{Bond:1997wr}.  The CMB intensity
spectrum has been measured with high
precision on large angular scales ($\ell < 600$) by the WMAP
experiment \citep{Hinshaw:2003ex}, while smaller angular scales have been probed
by ground and balloon-based CMB experiments \citep{Ruhl:2002cz, Readhead:2004gy,
Dickinson:2004yr, Kuo:2002ua, Halverson:2001yy}.  These data are broadly consistent with a
$\Lambda$CDM model in which the Universe is spatially flat and is
composed of radiation, baryons, neutrinos and the ever mysterious duo, cold dark
matter and dark energy.  

One of the firm predictions of this {\it standard model} is that the
CMB is intrinsically polarized.
Observations of the polarization
power spectra, and the correlation with the total intensity spectra can therefore
be used as a powerful consistency check, as well as potentially
providing additional cosmological information. On large angular scales
the polarization is sensitive to the details of the reionization
history and the curl component is a unique signature of tensor
perturbations.  On smaller angular scales the polarization spectra can
verify some of the basic assumptions made in the standard model.
For instance, peaks in the polarization spectra arise from
the same acoustic oscillations at last scattering as those in the total intensity
spectra.  However, the peaks in the polarization spectra are predicted to be
out of phase with the intensity peaks since the former are sourced by the velocity
term of the photon-baryon fluid as opposed to its density.
This effect provides the strongest constraint on the origin of the structure
observed in the spectra and breaks the severe degeneracy that is
introduced in models with radically broken scale invariance.  These are
models in which non-trivial structure may already exist in the spectrum of initial
perturbations.  

The recent polarization measurements of the DASI
\citep{Kovac:2002fg}, CAPMAP~\citep{Hedman:2002ck}, WMAP \citep{Kogut:2003et}, and CBI \citep{Readhead:2004xg} experiments have
confirmed that
the CMB is indeed polarized, providing an independent means for testing the underlying
model.  Many of the standard model cosmological parameters are becoming
highly constrained, especially in combination with complementary data sets \citep[e.g.][]{Seljak:2004xh}.

In this paper we test the standard model against the data from the 2003 
long-duration balloon (LDB) flight of the BOOMERANG experiment (hereafter
B03).  This mission marks the instrument's second successful
trip over the Antarctic continent.
The first LDB flight in December of 1998 (hereafter B98), resulted in
landmark, high signal-to-noise maps of the CMB intensity anisotropy
and the detection of the first few peaks of the intensity angular power
spectrum \citep{deBernardis:2000gy,Netterfield:2001yq, Ruhl:2002cz}.  For the 2003 flight
the instrument receiver was redesigned to be polarization sensitive and
the pointing system was upgraded to enable better attitude
reconstruction.  The \BOOMERANG\ sky coverage is
comprised of $\sim$195 hours of data over $\sim$1.8\% of the sky, with
an effective beam 11.5 $\pm$ 0.23 arcminutes.  Instrument
calibration is based on the 90 GHz and 60 GHz WMAP data and the
resulting amplitude uncertainty in calibration is $\sim2$\%.
A complete instrument description, and the \BOOMERANG\ CMB and Galactic maps
are given in \cite{Masi:2005}.  The
final data set from the flight is comprised of four power spectra:  the intensity power
spectrum, TT; the EE (curl-free) and BB (curl-like) polarization
power spectra; and the TE cross-power spectrum.  These spectral data are
presented in \cite{Jones:2005}, \cite{Montroy:2005} and \cite{Piacentini:2005}.

This analysis examines in detail the cosmological implications of the
\BOOMERANG\ data set.  We begin by outlining our required data products and methodology in
Section~\ref{method}.  We describe in Section~\ref{combos} the various
data combinations that are used in this analysis.  
In Section~\ref{baseline} we focus on the standard $\Lambda$CDM model and,
applying only weakly restrictive priors,
we find that the simple parameter fits to \BOOMERANG\ data alone are
fully consistent with those derived from other existing CMB data.  To this CMB
data, including the B03 data, we add in recent
Large Scale Structure (LSS) redshift survey data, consisting of matter power spectra
from the Sloan Digital Sky Survey (SDSS) \citep{Tegmark:2003uf} and the 2 Degree
Field Galaxy Redshift Survey (2dFGRS) \citep{Percival:2001hw}, and determine the marginalized
parameter constraints from this combined cosmological data set.  In Section~\ref{modified} we
extend our analysis to include tests of
several modifications of the standard model with the combined data
sets.

All of the models in Section~\ref{adiabatic} share the assumption that
the initial perturbations of the primordial plasma are adiabatic: in the
early radiation dominated era the matter and radiation densities are all
identically perturbed, giving an overall total density and hence curvature
perturbation.  This is not, however, the only possibility.  Isocurvature
modes describe the other linear combinations of matter and radiation
perturbations that do not contribute a curvature perturbation initially.
Models with isocurvature contributions to the perturbations give rise to
distinct signatures in the total intensity and polarization spectra.
The latter can be used to further constrain the possible contributions by
isocurvature modes that are not ruled out by measurements of the total
intensity spectrum alone.  In Section~\ref{iso} we explore the
constraints of the \BOOMERANG\ and other data on a model with a mixture
of a dominant adiabatic mode and a sub-dominant isocurvature mode.

\section{Data Products and Methodology}
\label{method}
\subsection{Summary of \BOOMERANG\ Results}
\label{xfaster}

We have developed two parallel and independent pipelines that we use to reduce
the \BOOMERANG\ observations from the
time-ordered data to polarization and intensity anisotropy maps,
through to angular power spectra. One pipeline was developed
predominantly in North America (NA pipeline)
\citep{Contaldi:2005,Jones:2005a} and the other predominantly in Italy
(IT pipeline)\citep{Masi:2005}. The purpose of constructing two
separate end-to-end pipelines is to check for self-consistency at
various stages during the reduction and to check for robustness in the
final spectra. We have carried out an extensive comparison of the
output of the NA and IT pipelines. We find excellent agreement for both
the TT spectrum \citep{Jones:2005}, representing the high
signal-to-noise limit, and the polarization spectra
\citep{Montroy:2005, Piacentini:2005} which 
are the most sensitive to the treatment of the experiment's noise characteristics
and systematics.  \BOOMERANG\ spectra obtained from the IT pipeline were                        
tested with the same weak priors for the standard model case.  The            
resulting parameter determinations are in good agreement with those           
reported here.

The parameter constraints presented in this analysis are based on the output of
the \XFASTER\ hybrid Monte Carlo--maximum likelihood estimator
\citep{Contaldi:2005}\footnote{\url{http://cmb.phys.cwru.edu/boomerang} \\ \url{http://oberon.roma1.infn.it/boomerang/b2k} }. The
estimator uses a close to optimal, quadratic, Fisher matrix based
estimator which is calibrated using signal only and noise only
simulations of the entire data set, from time stream to final maps. It
determines true polarization and total intensity angular power spectra
(averaged over pre-determined $\ell$ bands) on the sky.  After an
arbitrary initial guess, the quadratic estimator iterates onto the
maximum likelihood solution \citep{Bond:1998qg}, $ C_{B}^{\rm dat} $, with
errors determined by an estimate of the Fisher matrix for all
band powers self-consistently.  This ensures that the variance for each
band power includes contributions from all cross terms and from all spectra. This is
particularly important in the case of the cross spectra, as for example with the TE sample variance which is susceptible to the
TT and EE power in addition to the TE power itself. 

The calculation of the full Fisher matrix also allows us to exclude band
powers self consistently by cutting rows and columns from the inverse
Fisher matrix.  The effect of reduced sky coverage
and/or pixel weighting is accounted for by computing all coupling
kernels following \cite{Hivon:2001jp} and \cite{Chon:2003gx}. The analysis
typically includes a simultaneous determination of a complete set of
TT, EE, BB, TE, TB, and EB band powers. The EB and TB spectra are
consistent with zero (as expected) and are excluded from the parameter
determination by cutting out the bands in the inverse Fisher matrix
(equivalent to marginalizing over their contribution).

\begin{figure*}[!t]
\begin{center}
\rotatebox{270}{\resizebox{5in}{!}{
  \includegraphics*{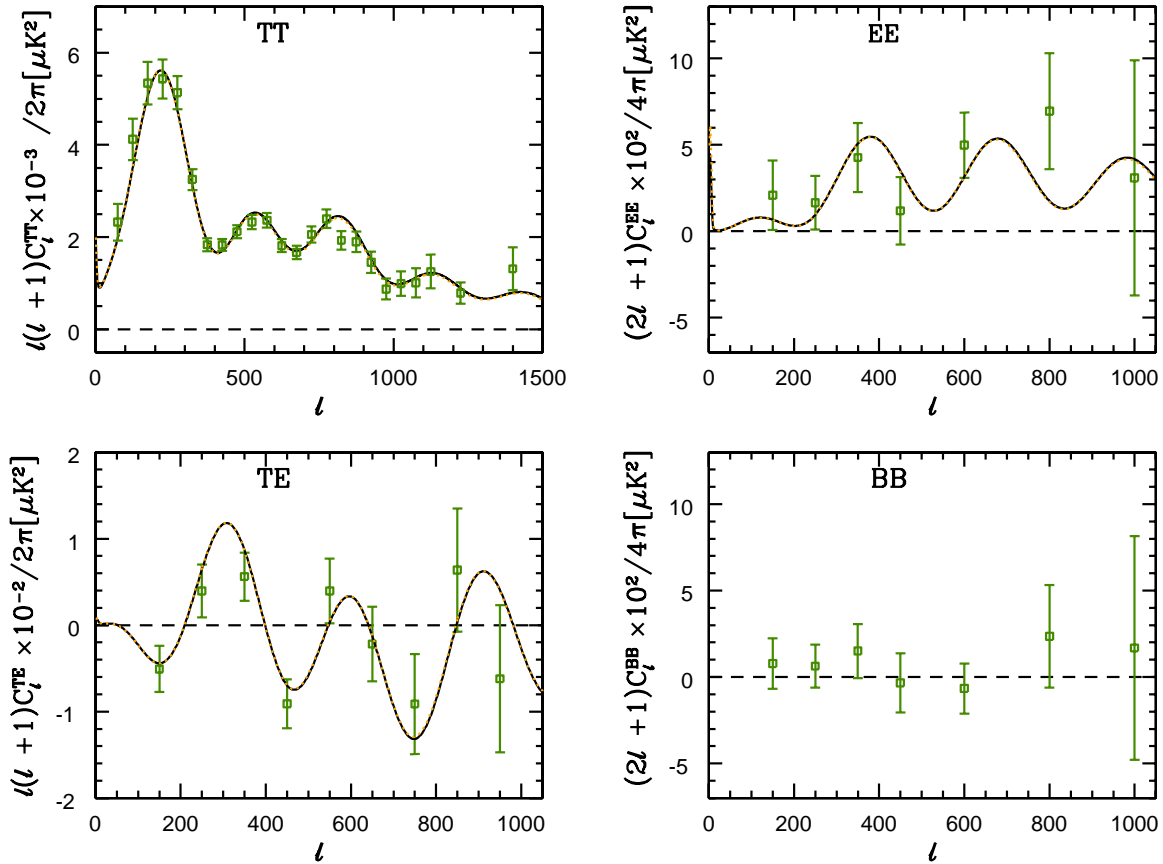}
}
}
\end{center}
\caption[b2kdata]{\small The \BOOMERANG\ bandpowers used in this
  analysis. We have included the total intensity TT, polarization
  EE and BB, and cross correlation TE spectra. The EB and TB spectra
  are excluded from this parameter analysis.  The solid/black curve is
the previous concordance model, a best fit to WMAP(first-year)+CBI+ACBAR
data from \url{http://lambda.gsfc.nasa.gov/product/map/}, with ($\Omega_b
h^2, \Omega_c h^2, n_{\rm s}(k=0.05), \exp(-2\tau), A(k=0.05), h$) =
(0.0224, 0.111, 0.958, 0.802, 0.739, 0.720).
The yellow/dotted curve is the CMBall (Table~\ref{tab:cmball})+B03 maximum
likelihood $\Lambda$CDM model from this analysis with (slightly
different parameterization--see text), ($\Omega_b h^2, \Omega_c h^2,
n_{\rm s}(k=0.05),
\tau, \ln(10^{10} A_{\rm s}(k=0.05)), \theta$) = (0.0228, 0.108, 0.959, 0.138,
3.12, 1.04).  
\label{fig:b2kdata}}
\end{figure*}

The spectra used in this
analysis are shown in Figure~\ref{fig:b2kdata}. The data have been
divided into bands which are generally $\Delta\ell$ = 50 wide for TT
and $\Delta\ell$ = 100 wide for the three remaining spectra.  The multipole
ranges for the B03 spectra which are used in this analysis are presented
in Table~\ref{tab:B03data}.  All band-to-band correlations are
included in the Fisher information matrix and are at most
$\sim$20\%. The band spacing was chosen in part to ensure that these
correlations were not large.

\begin{table*}[h]
\caption{\rm B03 bandpowers. The lowest bandpowers of the TT spectrum           
($\ell < 375$) are excluded when combining the \BOOMERANG\ data with        
the WMAP results since the two spectra are signal dominated and             
                        therefore correlated.   \label{tab:B03data}}
\space
\begin{tabular}{|c|c|c|c|}
\hline\hline
& & & \\
{\bf \BOOMERANG\ Spectrum} & {\bf Multipole Range} & {\bf Number of Bands} & {\bf Reference}\\
& & & \\
\hline
\space
& & & \\
TT  &  75 (375) $\leq \ell \leq 1400 $    & 24 (18) &\cite{Jones:2005}  \\
TE  & $ 150 \leq \ell \leq 950 $    & 9 &\cite{Piacentini:2005}   \\
EE \& BB   & $ 150 \leq \ell \leq 1000 $   & 7 & \cite{Montroy:2005}  \\
\hline\hline
\end{tabular}
\centering
\end{table*}

The \XFASTER\ code also calculates the required band window functions,
$W_{\ell}^{B}$, which are used to convert the model power spectra,
$C^{\rm mod}_{\ell}$, into theoretical bandpowers via
\begin{equation}
\langle C^{\rm mod}_{B} \rangle  = \frac{{\cal I}[
W^{B}_{\ell}{\cal C}^{\rm mod}_{\ell}]}{{\cal I}[W^{B}_{\ell}]}
\end{equation}
Here ${\cal C}^{\rm mod}_{\ell} = \ell(\ell + 1)C^{\rm mod}_{\ell}/2\pi $ and
we have introduced the notation for the ``logarithmic integral'' of a
spectrum \citep{Bond:1998qg}, ${\cal I}[f_{\ell}] \equiv \sum_{\ell}
\frac{\ell + \frac{1}{2}}{\ell (\ell + 1)}f_{\ell} $.  The above
operation permits direct comparison of theory $C^{\rm mod}_{B}$ with data
$ C_{B}^{\rm dat} $.

A final issue is the potential bias introduced by the
non-Gaussian distribution of the bandpowers in the signal dominated
regime. It has been shown
\citep{Bond:1998qg} that the variable $Z_{B} = \ln(C_{B}^{\rm dat} +
C_{B}^{N})$ is more normally distributed than the bandpowers $C_B$.
The noise offsets, $C_{B}^{N}$, are a measure of the deconvolved noise
spectrum on the sky and are calculated with the same quadratic
estimator using \XFASTER\ on the average of simulated 
noise-only observations. 

The distribution of the bandpowers tends to a Gaussian in the noise
dominated regime and log-normal in the sample variance dominated
regime. Both limits are significant for the TT bandpowers, hence we
transform all the TT bands to offset log-normal variables and treat the
likelihood function in the new variables as Gaussian for parameter
estimation.  For the polarization spectra EE and BB, which are
noise-dominated, we use $Z_B = C_{B}^{\rm dat}$, with no non-Gaussian
correction. For TE we also use $Z_B= C_{B}^{\rm dat}$ since negative
values of $ C_{B}^{\rm dat}$ occur.
The Fisher matrix of the
bandpowers is transformed as ${\widetilde F}_{BB'} = Z_B^\prime F_{BB'} Z_{B'}^\prime$
with $Z_B^\prime \equiv dZ_B / dC_B^{\rm dat} = (C^{\rm dat}_B + C_B^N)^{-1}$ if $B$ is a TT bandpower and
$Z_B^\prime = 1$ otherwise.

In
summary the \XFASTER\ data products include the bandpowers, Fisher
matrix, window functions and noise offsets.

\subsection{Parameter Estimation Methodology}
\label{extraction}

The Monte Carlo Markov Chain (MCMC) sampling technique we use for parameter estimation is
described in detail in \cite{Neal93,Christensen:2001gj,Lewis:2002ah}
and implemented in the publicly available
\COSMOMC\footnote{\url{http://cosmologist.info/cosmomc}} package.
Here we give a brief summary of the relevant details.  The technique
uses a Bayesian approach, generating samples of the posterior
probability density function (PDF) of the parameters 
${\bf y}$ given the data ${\bf z}$:
\begin{equation}
P( \mbox{\boldmath $y$}|{\bf z}) \propto P(\mbox{\boldmath
$y$})P({\bf z}|\mbox{\boldmath $y$}),
\end{equation}
where \( P({\bf z}|\mbox{\boldmath $y$}) \) is the likelihood PDF
and \( P(\mbox{\boldmath $y$}) \) is the prior PDF of \(
\mbox{\boldmath $y$} \).  The posterior is sampled by running a
number of Markov Chains.  The chains are constructed via the
Metropolis-Hastings (MH) algorithm whereby a candidate parameter
vector, \( \mbox{\boldmath $y^\prime$} \) is determined from an
arbitrary {\it proposal density} distribution \(
q(\mbox{\boldmath$y^\prime$}|\mbox{\boldmath $y_{n}$}) \) where
\( \mbox{\boldmath $y_{n}$} \) is the current state of the chain.
The candidate \( \mbox{\boldmath $y^\prime$} \) is accepted with {\it
acceptance probability} given by
\begin{eqnarray}
\alpha(\mbox{\boldmath $y$}|\mbox{\boldmath $y_{n}$}) =
\mbox{min} \left\{ \frac { P(\mbox{\boldmath $y^\prime$}|{\bf
z})q(\mbox{\boldmath$y_{n}$}|\mbox{\boldmath $y^\prime$})} {
P(\mbox{\boldmath $y_{n}$}|{\bf
z})q(\mbox{\boldmath$y^\prime$}|\mbox{\boldmath $y_{n}$})}, 1 \right\}.
\end{eqnarray}
At each point in the chain the acceptance probability for a candidate
point is compared to a random number $u$ drawn uniformly in the 0 to 1
range.  If \( u \leq \alpha(\mbox{\boldmath $y^\prime$}|\mbox{\boldmath
$y_{n}$}) \) then the proposed vector is accepted and the next point in
the chain is \( \mbox{\boldmath $y_{n+1}$} \) = \( \mbox{\boldmath
$y^\prime$} \).  If \( u > \alpha(\mbox{\boldmath
$y^\prime$}|\mbox{\boldmath $y_{n}$}) \) then the proposed vector is
rejected and \( \mbox{\boldmath $y_{n+1}$} \) = \( \mbox{\boldmath
$y_{n}$} \).

For the \BOOMERANG\ CMB data the likelihood evaluation at each point in the chain
requires the calculation of
\begin{equation}
\chi^{2} = \sum_{BB'}(Z_B^{\rm mod}(\mbox{\boldmath $y$}) -
Z_{B}^{\rm dat}){\widetilde F}_{BB'}(Z_{B'}^{\rm mod}(\mbox{\boldmath $y$})
- Z_{B'}^{\rm dat}).
\end{equation}
The WMAP data likelihood is computed using the likelihood code supplied
by the WMAP team~\citep{Verde:2003ey,Kogut:2003et}, but with two
modifications.  The first modification is a change to the TE likelihood
function to account for the correlation between the intensity and TE
power spectrum estimators (we neglect the small correlations between the
$C_l$ estimators at different $l$)~\citep{Dore:2003wp}.  After the
chains have been run we use importance sampling
(e.g. see~\citet{Lewis:2002ah}) to correct the WMAP likelihood on large
scales using the more computationally intensive likelihood code
from~\citet{Slosar:2004fr}. This {\it Slosar-Seljak modification} uses a more accurate
calculation of the WMAP likelihood at low multipoles ($\ell \leq 11$)
and considers in more detail the errors associated with foreground removal.

The theoretical CMB spectra (as well as the matter power spectra) are
computed using CAMB~\citep{Lewis:1999bs}, a fast parallel Boltzmann code
based on CMBFAST~\citep{Seljak96}.  We calculate statistics of interest,
such as the marginalized posterior distribution of individual
parameters, from the MCMC samples after removing burn in. We run six
chains for each combination of data and parameters that cannot be
importance sampled.  We marginalize numerically over each data point's calibration
and beam uncertainties at each sample in the chain. The calibration errors are assumed to be
independent between data sets.  
We check convergence by ensuring that the standard
deviation between chains of the $95\%$-percentile estimated from each
chain is less than 0.2 in units of the all-chain parameter standard
deviation. This should ensure that sampling errors on quoted limits
are minimal.

Parameter estimates from MCMC have been shown to be in very good
agreement with those derived using an adaptive ${\cal C}_\ell$-grid
\citep{Bond:2003ur} that was previously applied to the B98
analysis \citep{Ruhl:2002cz}.  MCMC results for CMBall+B98
\citep{Bond:2004rt} are also in good agreement with those
we obtain for CMBall+B03 for the baseline model defined below with
the same priors applied\footnote{\BOOMERANG\ and B98, with overlapping sky coverage, are correlated data sets.  We therefore exclude B98 from   
this analysis and will consider the combined B98 and \BOOMERANG\ maps in a future analysis.}.

\section{Data Combinations}
\label{combos}
\subsection{The CMB data}

We consider a number of combinations of data.                                 
We break the \BOOMERANG\ data set into one subset consisting of the                                   
TT spectrum alone (\BOOMERANG TT), and another subset consisting                                      
of the EE, BB and TE spectra (\BOOMERANG pol) alone.  We also consider                                
fits to the entire \BOOMERANG\ data set, WMAP data alone and a                                        
combined \BOOMERANG\ + WMAP data set.  We next combine \BOOMERANG\                                    
with available data from a collection of CMB experiments.  We outline                                 
in Table~\ref{tab:cmball} the experiments and multipole ranges which                                  
make up that collection, which we call CMBall.  We note that because of the overlap in $\ell$                                 
range of the ARCHEOPS \citep{Tristram:2004ke} data with the WMAP data,                                
the former cannot be included in the CBMall data set, unless a joint
analysis is done.    
The \BOOMERANG\ multipole range is given in                                    
Table~\ref{tab:B03data}.  The cosmic variance of the WMAP and                  
\BOOMERANG\ data sets is correlated in the low multipole range                 
(essentially over the first peak of the TT power spectrum).  To account for    
this, we cut the lower multipoles of the B03 TT                                
spectrum ($\ell < 375$) when combining \BOOMERANG\ data with WMAP              
data. 

\subsection{The LSS Data}

For our final data combination                                                                     
we also include LSS observations from                                                              
2dFGRS and the SDSS.  The two redshift surveys are treated in a                                    
conservative fashion. For example, although the SDSS bandpowers have                               
been corrected for differing galaxy bias factors associated with                                   
different types of galaxies, there is still an overall galaxy bias                                 
factor, $b_g$, the ratio of the square root of the galaxy-galaxy power                             
spectrum for $L_*$ galaxies to that of the mass                                                    
density power spectrum today. Although the indications are that this is a number near unity                             
\citep{ Percival:2001hw, Tegmark:2003uf}, in our standard results we allow it to take on            
arbitrary values by marginalizing it over a very broad distribution. This                          
means that our LSS information is only constraining models through the shape                       
of the power spectrum, but not the overall amplitude. Constraining the                             
overall amplitude is akin to imposing a prior on $\sigma_8$. To test                               
sensitivity to this, we have adopted varied Gaussian errors on $b_g^2$                             
about a mean.  We have taken the mean to be unity and adopted errors on                            
$\delta b_g^2$ appropriate for $\delta b_g = 0, 0.1, 0.5$ and $10$,                                
then marginalized over $b_g^2$. A uniform prior in $b_g^2$ leads to                                
the same results as for $\delta b_g = 10$  
\footnote{Note that allowing $b_g^2$ to be
negative has no effect and yields the same results as a (uniform)
positive $b_g^2$ constraint.}.  Most parameter averages we                              
obtain are relatively insensitive to $\delta b_g $. We comment on its                              
effect below: it has impact on the massive neutrino and dark energy                                
equation of state constraints. We only use SDSS data for wavenumbers                               
$k < 0.1 h {\rm Mpc}^{-1}$ to avoid nonlinear corrections and to avoid                             
possible non-uniform $b_g$ complications. (See \cite{Tegmark:2003uf}                               
for a discussion of these and other issues.) A similar $\delta b_g$                                
marginalization strategy was used for the 2dFGRS redshift survey data.

An estimate using galaxy-galaxy lensing from SDSS \citep{Seljak:2004sj}
is $b_g =0.99 \pm 0.07$. (These authors also used WMAP data to obtain
this value, so it is not a completely independent determination of the
bias.)  An estimate
using the 3-point function and redshift space clustering distortions
for 2dFGRS gives $b_g =1.04 \pm 0.04$ \citep{Verde:2001sf}. Based on these
two analyses, adopting $b_g =1.0 \pm 0.10$ to illustrate the effect of
knowing the bias better, which translates into a $\sigma_8$ prior,
seems reasonable.

For the purposes of this paper, in which our focus is on the B03 CMB data,
we have limited the LSS information we include. For example, we
have not incorporated the SDSS results on luminous red galaxies
\citep{Eisenstein:2005su}. The recent final power spectrum and window
functions of the 2dFGRS survey \cite{Cole:2005sx} is not yet available.

\subsection{Other Data Sets}

We have applied the supernova data (SNIa) in Section~\ref{darke} to the determine            
the dark energy equation of state.  For                                        
this we use the {\it gold} set, as described in \cite{Riess:2004nr}.                                     
Also, for a few cases we include the $H_0$ prior value from the HST Key
Project ~\citep{Freedman:2000cf}.

We do not explicitly include weak lensing results.  These generally determine
the parameter combination $\sigma_8 \Omega_m^{0.8} $ providing
additional independent constraints \citep{Contaldi:2003hi,Bond:2002tp,Readhead:2004xg}.
We also do not include information on the Lyman alpha forest, even
though it probes the power spectrum to smaller scales. Although adding
this data does result in some more stringent constraints than those we
derive here \citep{McDonald:2004eu}, the forest information is more susceptible to scale
dependent biasing effects associated with gasdynamical and radiation
processes.

\section{Adiabatic Models}
\label{adiabatic}
\subsection{Baseline Model}
\label{baseline}

\subsubsection{Parameterization and Priors}
For our baseline model we consider a flat universe with photons,
baryons, massless neutrinos, cold dark matter and a cosmological
constant.  Initial conditions will be taken to be purely adiabatic (no
isocurvature modes).  We assume a power law form for the power
spectrum of the primordial comoving curvature perturbation,
described by ${\cal P}_{\rm s} = A_{\rm s}(k/k_\star)^{(n_{\rm s}-1)}$,
where the $n_{s}$ is the scalar spectral spectral index and
$A_{s}$ is the scalar amplitude (we choose a pivot point $k_\star =
0.05\Mpc^{-1}$). The physical baryon density and dark matter
density are parameterized by $\Omega_{b}h^{2}$ and $\Omega_{c}h^{2}$,
where $h = H_0/100 {\rm km}$ ${\rm s}^{-1}\Mpc^{-1}$ is the
Hubble parameter.  We use the parameter
$\theta$ to characterize the positions of the peaks in the angular
power spectra, defined as one hundred times the ratio of the sound
horizon to the angular diameter distance at last
scattering~\citep{Kosowsky02}.  Finally, the parameter $\tau$ is used
to describe the Thomson scattering optical depth to decoupling.  Thus
our baseline model is a function of 6 cosmological parameters to which
we impose the following flat weak priors:                                  
$0.5 \leq n_{s} \leq 1.5$; $2.7 \leq \ln(10^{10} A_{s})       
\leq 4.0$; $0.005 \leq \Omega_{b}h^{2} \leq 0.1$; $0.01 \leq  
\Omega_{c}h^{2} \leq 0.99$; $0.5 \leq \theta \leq 10.0$; and  
$0.01 \leq \tau \leq 0.8$.     
Additional weak priors restrict the age of the universe to $10 \Gyr
\leq \text{age} \leq 20 \Gyr$ and the expansion rate to $0.45 \leq h
\leq 0.9$.  All priors are summarized in Table~\ref{tab:weakpriors}.
Besides being generally agreed upon by cosmologists, our
weak priors are consistent with those used in much of the CMB
literature, {\it e.g.,} \cite{Lange:2000iq}, \cite{Bond:2003ur} and \cite{Readhead:2004xg}.   
We choose not to impose the restrictive
prior $\tau<0.3$ applied in \cite{Spergel:2003cb}.  We note that some of our results
are sensitive to our choice of prior on $H_0$ and we explore the effect of
strengthening our $H_0$ prior in Sections ~\ref{run} and ~\ref{curve}.

\begin{table*}[h]
\centering
\caption{\rm List of weak priors imposed on baseline parameter set.
Priors are uniform in the variable shown.\label{tab:weakpriors}  }
\space
\begin{tabular}{|c|c||c|c|}

\hline\hline 
& & & \\
{\bf Parameter} & {\bf Limits} & {\bf Parameter} & {\bf Limits} \\
& & & \\
\hline
& & & \\
$ \Omega_b h^2 $& 0.005 - 0.1 & $ n_{\rm s} $& 0.5 - 1.5\\
$ \Omega_c h^2 $& 0.01 - 0.99 & $ \ln[10^{10} A_{\rm s}] $& 2.7 - 4.0 \\
$ \theta $& 0.5 - 10.0 & $ {\rm Age(Gyr)} $ & 10 - 20 \\
$ \tau $& 0.01 - 0.8 & $ H_0 $& 45 - 90 \\ 
\hline\hline
\end{tabular}

\centering 
\end{table*}

In addition to the base parameter values, the results also include marginalized             
constraints for several derived parameters including:                                      
$\Omega_{\Lambda}$, the relative dark energy density; the age of the Universe;             
$\Omega_m$, the relative total matter density; $\sigma_8$, the root mean square            
linear mass perturbation in $8h^{-1}$ Mpc spheres; $z_{re}$, the                           
redshift of reionization assuming it is a sharp transition; and the
Hubble constant, $H_0$.  

\subsubsection{Consistency of \BOOMERANG\ Data Set}

\begin{figure*}[!t]
\begin{center}
\resizebox{6in}{!}{\includegraphics*{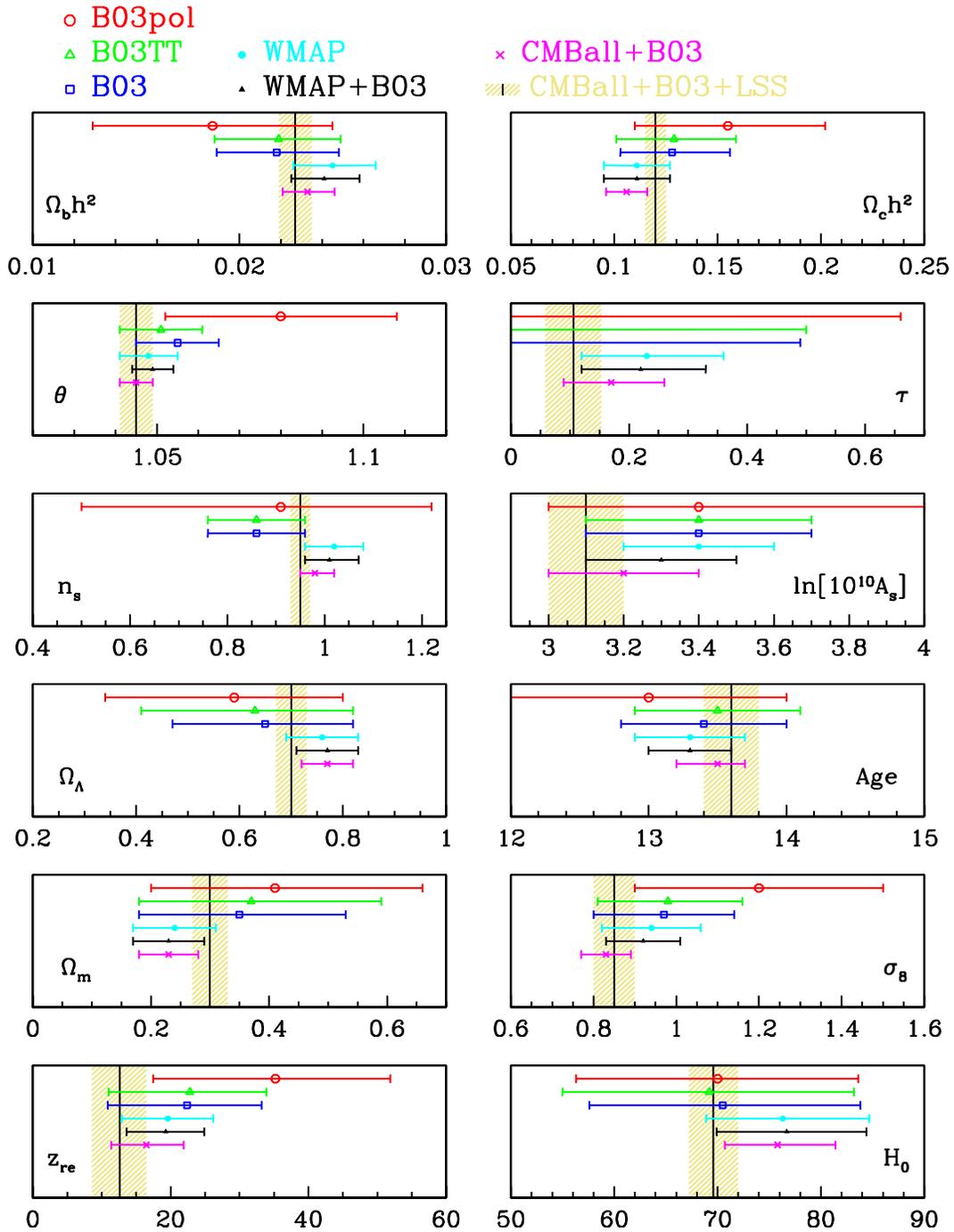}}
\end{center}
\caption[b2kdata]{\rm
Median values obtained from the marginalized
probability for each parameter for the baseline, standard model.  The
errors bars represent the 68\% confidence interval.  The  95\%
upper limit is given for the case of $\tau$ for \BOOMERANG\ data alone.
The following flat weak priors are imposed (as outlined in
Table~\ref{tab:weakpriors}):  
$0.5 \leq n_{s} \leq 1.5$; $2.7 \leq \ln(10^{10} A_{s})
\leq 4.0$; $0.005 \leq \Omega_{b}h^{2} \leq 0.1$; $0.01 \leq
\Omega_{c}h^{2} \leq 0.99$; $0.5 \leq \theta \leq 10.0$; and
$0.01 \leq \tau \leq 0.8$.  Additional weak priors restrict
the age of the universe to $10 \Gyr \leq \text{age} \leq 20 \Gyr$ and the
expansion rate to $45 \leq H_0 \leq 90$.  
Our baseline CMBall+B03+LSS result is fairly insensitive to $\delta b_g$        
and we have chosen for this case the less restrictive flat, uniform
prior in $b_g^2$.  

\label{fig:mean}}
\end{figure*}

The resulting marginalized parameter constraints for the baseline
model for each of the data combinations are given in
Table~\ref{tab:base} and presented graphically in
Figure~\ref{fig:mean}.  In both Table~\ref{tab:base} (and in the ones
that follow) and Figure~\ref{fig:mean} we give the Bayesian 50\%
probability value (the median) obtained from the marginalized
probability for each parameter.  The quoted errors represent the 68\%
confidence interval obtained by integrating the marginalized
distributions.  In the case of upper or lower bounds, the 95\%
confidence limits are quoted.  We note that our baseline
CMBall+B03+LSS result is fairly insensitive to $\delta b_g$ and that
we have chosen the less restrictive flat, uniform prior in $b_g^2$.

\begin{figure*}[!t]                                                                                                  
\begin{center}                                                                                                       
\resizebox{6in}{!}{\includegraphics*{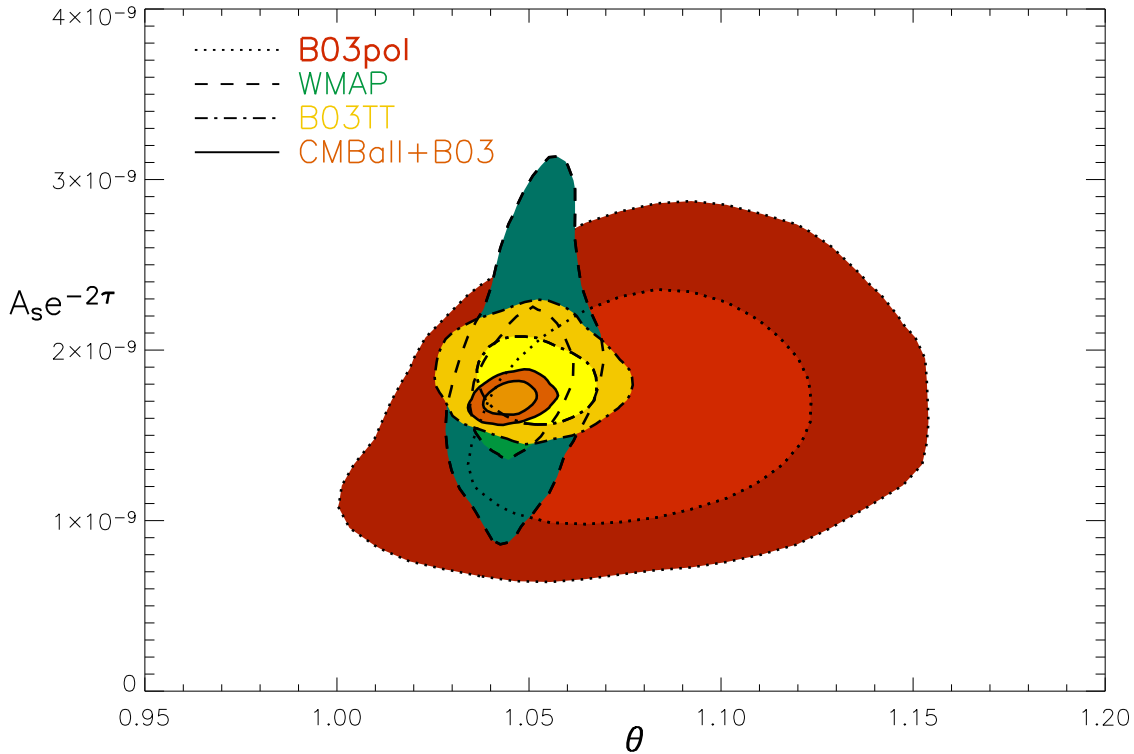}}                                                       
\end{center}                                                                                                         
\caption[b2kdata]{\small                                                                                             
Constraints on $A_{\rm s} e^{-2\tau}$ versus                                                                         
 $\theta$.  Inner contours represent 68\% likelihood regions and outer                                               
 contours 95\% likelihood regions.  The peak position characterization                                               
parameter $\theta$ is best the determined parameter in cosmology, $1.045                                             
\pm 0.004$, from the CMBall+\BOOMERANG\ data set.  We find that the \BOOMERANG TT data does particularly             
 well at constraining both the peak pattern and the combined  $A_{\rm s}                                             
e^{-2\tau}$ amplitude parameter.  The constraint from WMAP alone on                                                  
$A_{\rm s}$ is better than that from \BOOMERANG.  The                                                                
agreement between the \BOOMERANG pol and \BOOMERANG TT data is                                                       
consistent with the basic inflation picture.                                                                         
\label{fig:2Dlikecurves}}                                                                                            
                                                                                                        \end{figure*}

The comparison of \BOOMERANG pol and \BOOMERANG TT provides a robust
internal consistency check.  We note that the \BOOMERANG pol constraints
to $\Omega_{b}h^{2}$ and $\Omega_{c}h^{2}$ are quite good with
uncertainties which are only slightly larger than those of the \BOOMERANG TT result.
However, the \BOOMERANG pol constraints on $n_{\rm s}$, 
$\tau$ and $A_{\rm s}$ are weak and results for these cases are prior
driven.  We present in Figure~\ref{fig:2Dlikecurves} a 2D likelihood plot of                                  
$\theta$ versus the combined parameter $A_{\rm s} e^{-2\tau}$.  The latter                           
determines the overall power in the {\it observed} CMB anisotropy
(except at low $\ell$), and                               
is therefore better constrained than the primordial power $A_{\rm s}$.                               
\COSMOMC\ uses a covariance matrix for the parameters and is therefore                               
able to ascertain linear combination degeneracies. Although we use $\ln                              
A_{\rm s}$ and $\tau$ as base parameters, the proposal density knows that the                        
combination $\ln (A_{\rm s} e^{-2\tau})$ is well constrained                                         
and can explore the poorly constrained orthogonal direction                                          
efficiently. We find that the \BOOMERANG\ data alone does particularly                               
well at constraining $A_{\rm s} e^{-2\tau}$.  The angular-diameter distance                          
variable $\theta$ defines the shift                                                                  
with $\ell$ of the overall ${\cal C}_\ell$ pattern, in particular of the                             
pattern of peaks and troughs. With all of the CMB data it is the best                                
determined parameter in cosmology, $1.045 \pm 0.004$; with                                           
\BOOMERANG pol it is an important test which demonstrates the consistency of the positions of        
the polarization spectra peaks and troughs relative to those forecasted                              
from the TT data, although the errors are larger with $\theta = 1.08 \pm 0.03$.  For the 
CBI TT, TE and EE data in combination with WMAP                             
TT and TE, \citet{Readhead:2004xg} found $\theta = 1.044 \pm 0.005$. 
With just the CBI EE polarization data they determined $\theta = 1.06
\pm 0.04$, again showing the consistency we find of the data with the 
TT forecast of the polarization peaks and trough. 

\begin{figure*}[!t]
\begin{center}
\resizebox{6in}{!}{\includegraphics*{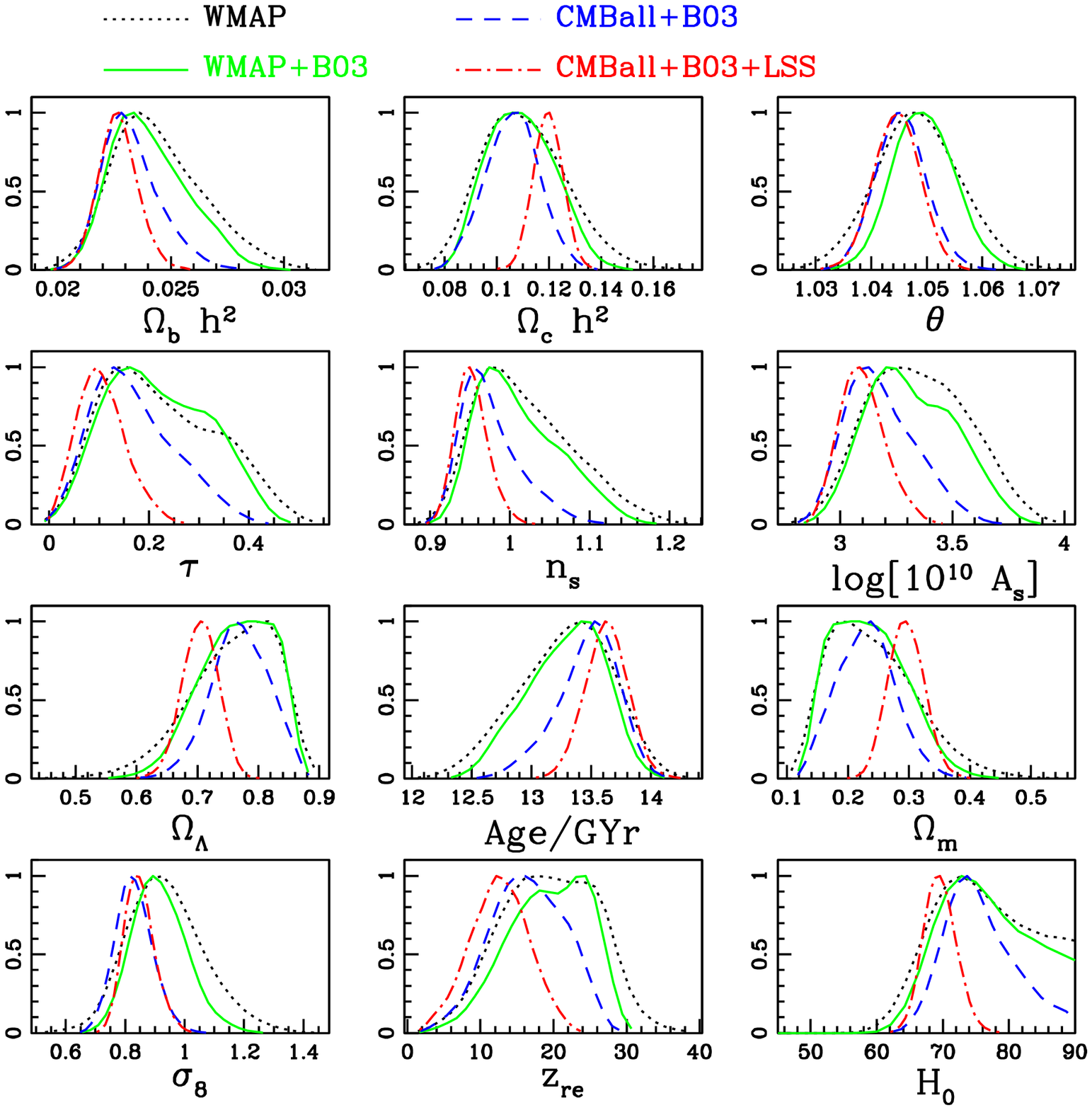}}
\end{center}
\caption[b2kdata]{\rm
Marginalized one-dimensional distributions for the baseline model parameters for the data
combinations WMAP only (black/dotted), WMAP + \BOOMERANG\ (green/solid), CMBall + \BOOMERANG\
(blue/dashed), and CMBall + \BOOMERANG\ + LSS (red/dash-dotted). The curves are each normalized by their
peak values. All distributions are derived from chains run with the weak set of external, uniform priors
shown in Table~\ref{tab:weakpriors}. The LSS data consists of the 2dFGRS
and SDSS redshift surveys (with a flat $b_g^2$ prior imposed).  The most significant impact of the B03 data
is on ${\sigma}_8$.  Moreover, the ${\sigma}_8$ constraint from CMB data alone
is quite strong, with the addition of LSS data having little effect.
\label{fig:likecurves}}
\end{figure*}

The \BOOMERANG\ median parameter values are remarkably consistent with the
parameter constraints from WMAP data alone.  We note that in general the
Slosar-Seljak modification to WMAP tends to broaden WMAP parameter likelihood
curves and that the most significant impact on the median values is in
$\tau$ ($\sim 0.3\sigma$ increase) and in $\Omega_m$ ($\sim 0.6\sigma$
decrease).  Adding \BOOMERANG\ to the WMAP data decreases the
parameter uncertainties by an average of 16\%.  The most significant
effect is a $\sim$33\% decrease in the $\sigma_8$ uncertainty.
Figure~\ref{fig:likecurves} shows the likelihood curves for the 6 base 
parameters and 6 derived parameters for a variety of
data combinations.  Overall the various data combinations are
generally in good agreement at better than the $1\sigma$ level.  The largest
outlier is $\Omega_c h^2$ which increases by $1.5\sigma$ with the
addition of the LSS data set.  Also, similar to \cite{Spergel:2003cb},
we find that the addition of small scale CMB data lowers both the
value for the amplitude of fluctuations at $k = 0.05 \Mpc^{-1}$ and the
value of the scalar spectral index.  The effect of adding the LSS data
follows this trend.

\begin{table*}
\centering
\caption{\rm The CMBall data set.\label{tab:cmball} }
\space
\begin{tabular}{|c|c|c|}
\hline\hline
& & \\
{\bf Experiment} & {\bf Multipole Range} & {\bf Reference} \\
& & \\
\hline
 &   &    \\
WMAP TT  & $ 2 \leq \ell \leq 899 $    &  \cite{Hinshaw:2003ex}   \\
WMAP TE  & $ 2 \leq \ell \leq 512 $    &  \cite{Kogut:2003et}   \\
DASI TT  & $ 380 \leq \ell \leq 800 $   &      \cite{Halverson:2001yy} \\
VSA TT   & $ 400 \leq \ell \leq 1400 $   &   \cite{Dickinson:2004yr}   \\
ACBAR TT & $ 400 \leq \ell \leq 1950 $   &     \cite{Kuo:2002ua} \\
MAXIMA TT & $ 450 \leq \ell \leq 1150 $ &     \cite{Hanany:2000qf} \\
CBI TT   & $ 750 \leq \ell \leq 1670 $   &   \cite{Readhead:2004gy}  \\
\hline\hline
\end{tabular}
  \end{table*}

\begin{table*}
\centering
\space
\caption{\rm 
Marginalized parameter constraints for the baseline, 6 parameter,
$\Lambda$CDM model.  Parameter uncertainties represent the 68\% confidence interval obtained by    
integrating the marginalized distributions.  95\% confidence limits are
quoted for the case of upper bounds.  
The following flat weak priors are imposed (as outlined in                                                              
Table~\ref{tab:weakpriors}):                                                                                            
$0.5 \leq n_{s} \leq 1.5$; $2.7 \leq \ln(10^{10} A_{s})                                                                 
\leq 4.0$; $0.005 \leq \Omega_{b}h^{2} \leq 0.1$; $0.01 \leq                                                            
\Omega_{c}h^{2} \leq 0.99$; $0.5 \leq \theta \leq 10.0$; and                                                            
$0.01 \leq \tau \leq 0.8$.  Additional weak priors restrict                                                             
the age of the universe to $10 \Gyr \leq \text{age} \leq 20 \Gyr$ and the                                               
expansion rate to $45 \leq H_0 \leq 90$.  The CMBall data set is as
given in Table~\ref{tab:cmball}.  The LSS data consists of the galaxy 
power spectra from the 2dFGRS and SDSS redshift surveys. Our                                                                                  
baseline CMBall+B03+LSS result is fairly insensitive to $\delta b_g$                 
and we have chosen for this case the less restrictive flat, uniform
prior in $b_g^2$.  The constraints from the \BOOMERANG pol data are in good
agreement with the \BOOMERANG TT data, although some parameters
constraints for the  \BOOMERANG pol case are prior driven, {\it eg.}
$n_{\rm s}, A_{\rm s}$ and $H_0$. \BOOMERANG\ does not constrain $\tau$,
but upper limits are given.  The constraints from the various data set are
consistent.
\label{tab:base} }
\scriptsize
\begin{tabular}{|c||c|c|c|c|c|c|c|}
\hline\hline
& & & & & & & \\
& B03pol & B03TT & B03 & WMAP & WMAP & CMBall & CMBall \\					
& & & & & +B03 & +B03 & +B03+LSS \\
& & & & & & & \\				
\hline
& & & & & & & \\
$       \Omega_b h^2 $ & $ 0.0187^{+0.0058}_{-0.0058} $  & $ 0.0219^{+0.0030}_{-0.0031} $ & $ 0.0218^{+0.0029}_{-0.0030} $    &$ 0.0245^{+0.0021}_{-0.0019} $ & $ 0.0241^{+0.0017}_{-0.0016} $ & $ 0.0233^{+0.0013}_{-0.0012} $   &  $ 0.0227^{+0.0008}_{-0.0008} $   	\\
$       \Omega_c h^2 $ & $ 0.155^{+0.047}_{-0.045} $     & $ 0.129^{+0.030}_{-0.028} $    & $ 0.128^{+0.029}_{-0.026} $       &$ 0.111^{+0.016}_{-0.016} $    & $ 0.109^{+0.013}_{-0.013} $    & $ 0.106^{+0.010}_{-0.010} $      &  $ 0.120^{+0.005}_{-0.005} $     	\\
$         \theta     $ & $ 1.080^{+0.028}_{-0.028} $     & $ 1.051^{+0.010}_{-0.010} $    & $ 1.055^{+0.010}_{-0.010} $       &$ 1.048^{+0.007}_{-0.007} $    & $ 1.049^{+0.005}_{-0.005} $    & $ 1.045^{+0.004}_{-0.004} $   	 &  $ 1.045^{+0.004}_{-0.004} $    	\\
$           \tau     $ & $ <0.66                $        & $    <0.50                $    & $  <0.49                 $       & $ 0.23^{+0.13}_{-0.11} $       & $ 0.22^{+0.11}_{-0.10} $       & $ 0.170^{+0.090}_{-0.081} $    	 &  $ 0.106^{+0.047}_{-0.048} $    	\\		    	
$         n_{\rm s}  $ & $ 0.91^{+0.59}_{-0.41} $        & $ 0.86^{+0.10}_{-0.10} $       & $ 0.86^{+0.10}_{-0.10} $          &$ 1.02^{+0.06}_{-0.06} $       & $ 1.01^{+0.06}_{-0.05} $       & $ 0.98^{+0.04}_{-0.03} $         &  $ 0.95^{+0.02}_{-0.02} $         	\\
$\ln[10^{10} A_{\rm s}]$&$  3.4^{+ 0.6}_{- 0.7} $        & $  3.4^{+ 0.6}_{- 0.2} $       & $  3.4^{+ 0.6}_{- 0.2} $          &$  3.4^{+ 0.2}_{- 0.2} $       & $  3.3^{+ 0.2}_{- 0.2} $       & $  3.2^{+ 0.2}_{- 0.2} $       	 &  $  3.1^{+ 0.1}_{- 0.1} $       	\\
$ \Omega_\Lambda     $ & $ 0.59^{+0.21}_{-0.25} $        & $ 0.63^{+0.19}_{-0.22} $       & $ 0.65^{+0.17}_{-0.19} $          &$ 0.76^{+0.07}_{-0.07} $       & $ 0.77^{+0.06}_{-0.06} $       & $ 0.77^{+0.05}_{-0.05} $       	 &  $ 0.70^{+0.03}_{-0.03} $       	\\
$     {\rm Age(Gyr)}     $ & $ 13.0^{+ 1.0}_{- 1.0} $        & $ 13.5^{+ 0.6}_{- 0.6} $       & $ 13.4^{+ 0.6}_{- 0.5} $          &$ 13.3^{+ 0.4}_{- 0.4} $       & $ 13.3^{+ 0.3}_{- 0.3} $       & $ 13.5^{+ 0.2}_{- 0.3} $       	 &  $ 13.6^{+ 0.2}_{- 0.2} $       	\\
$       \Omega_m     $ & $ 0.41^{+0.25}_{-0.21} $        & $ 0.37^{+0.22}_{-0.19} $       & $ 0.35^{+0.19}_{-0.17} $          &$ 0.24^{+0.07}_{-0.07} $       & $ 0.23^{+0.06}_{-0.06} $       & $ 0.23^{+0.05}_{-0.05} $       	 &  $ 0.30^{+0.03}_{-0.03} $       	\\
$       \sigma_8     $ & $  1.2^{+ 0.3}_{- 0.3} $        & $ 0.98^{+0.18}_{-0.17} $       & $ 0.97^{+0.18}_{-0.17} $          &$ 0.94^{+0.12}_{-0.12} $       & $ 0.92^{+0.09}_{-0.09} $       & $ 0.83^{+0.06}_{-0.06} $       	 &  $ 0.85^{+0.05}_{-0.05} $       	\\
$         z_{re}     $ & $ 35.2^{+16.7}_{-17.7} $        & $ 22.8^{+11.1}_{-11.8} $       & $ 22.0^{+11.3}_{-11.9} $          &$ 19.6^{+ 6.6}_{- 6.7} $       & $ 19.3^{+ 5.6}_{- 5.7} $       & $ 16.5^{+ 5.4}_{- 5.1} $       	 &  $ 12.6^{+ 3.9}_{- 4.0} $       	\\
$         H_0        $ & $ 70.0^{+20.0}_{-25.0} $        & $ 69.2^{+20.8}_{-24.2} $       & $ 70.4^{+19.6}_{-25.4} $          &$ 76.3^{+13.7}_{- 4.2} $       & $ 76.7^{+13.3}_{- 3.9} $       & $ 75.8^{+ 5.6}_{- 5.1} $      	 &  $ 69.6^{+ 2.4}_{- 2.4} $       	\\
\hline\hline
\end{tabular}
  \end{table*}

\subsection{Modified Standard Model}
\label{modified}

In this section we explore five extensions of the standard model by
adding, in turn, one parameter to the baseline parameter set.  In
all cases we maintain the same weak priors on the base parameters as
outlined in Table~\ref{tab:weakpriors}.  Some of the results                                                                 
are sensitive to our chosen prior range for $H_0$. For example, 
in certain cases we will note the impact of strengthening our $H_0$ prior
to the value from the HST Key Project ~\citep{Freedman:2000cf}, $h =
0.72 \pm 0.08 $, with the errors treated as Gaussian.  

\subsubsection{Running Index}
\label{run}
We modify the power law form for the power spectrum of the
density perturbations to allow the spectral index, $n_{\rm s}$, to vary with
scale.  Following \cite{Kosowsky:1995aa} this variation can be parameterized
by the term ${\nrun} = d n_{\rm s}/d\ln k$, such that $n_{\rm s} =
n_{\rm s}(k_\star) +
{\nrun}(k_\star)\ln(k/k_\star)$, where again $k_\star = 0.05\Mpc^{-1}$.  We restrict ${\nrun}$ to lie between -0.3 and 0.3.
Results from the combined data sets, CMBall+B03
and  CMBall+B03+LSS, are given in Table~\ref{tab:mod}.

\begin{figure*}
\begin{minipage}[t]{9cm}
\includegraphics[width=0.9\textwidth]{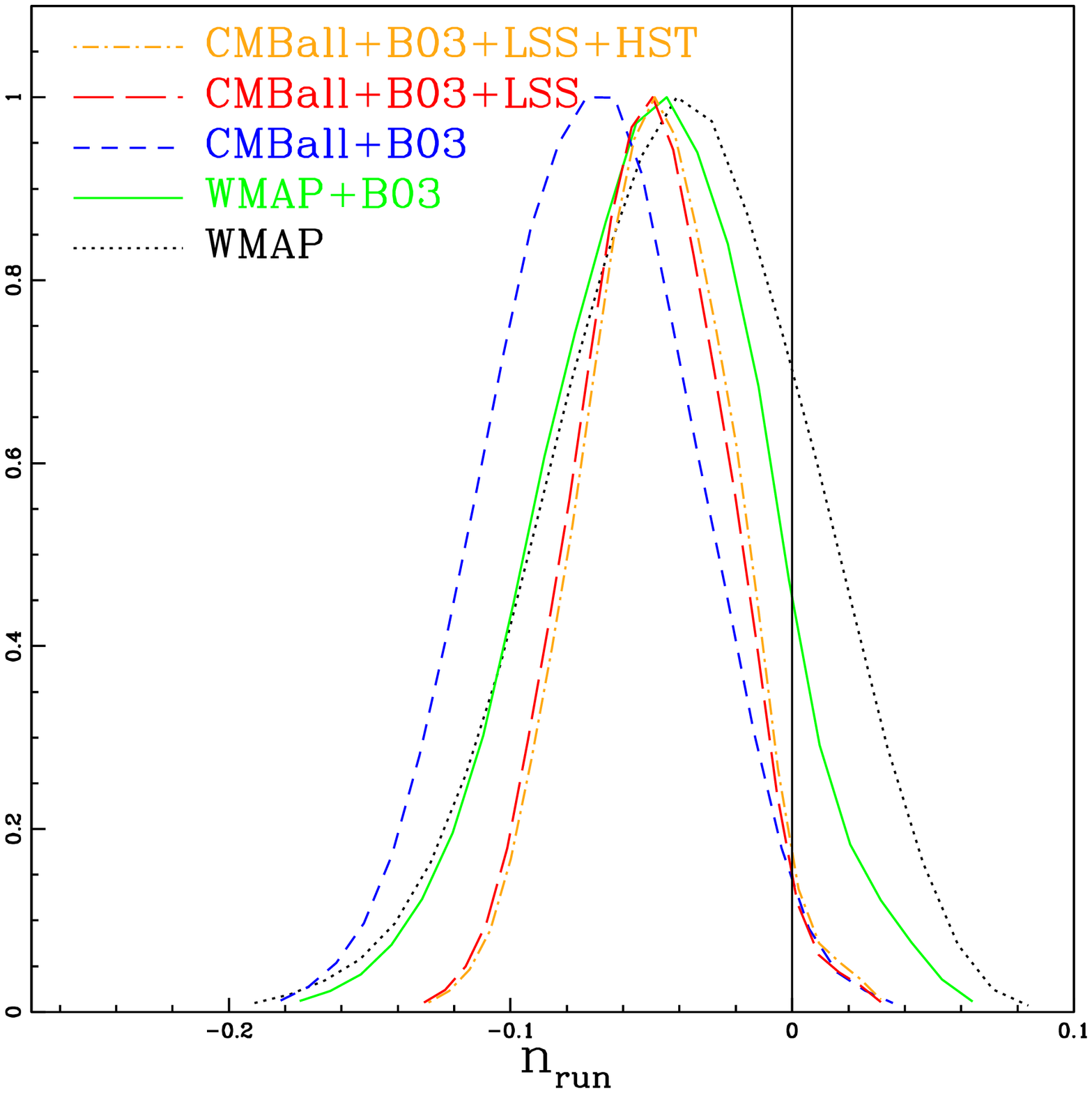}
\caption[b2kdata]{\small                    
Marginalized one-dimensional distributions for the ${\nrun}$ parameter for
the baseline + running index model.  Weak priors imposed are those outlined in
Table~\ref{tab:weakpriors}.  The running index parameter is restricted 
to lie between -0.3 and 0.3.  Application of the HST prior on $H_0$ 
slightly reduces the significance of a running index.
\label{fig:running}}

\end{minipage} 
\hfill
\begin{minipage}[t]{9cm}
\includegraphics[width=0.9\textwidth]{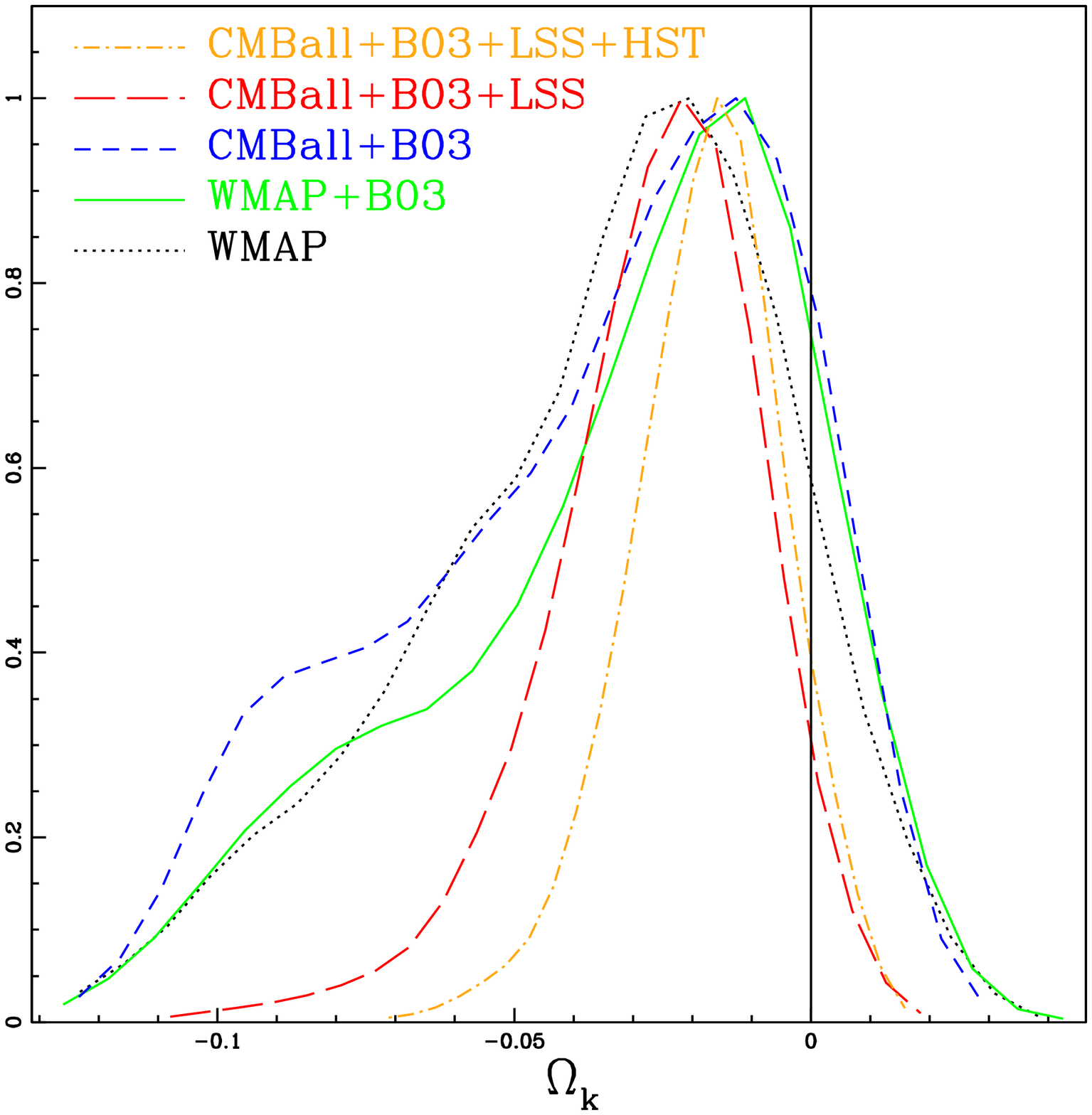}
\caption[b2kdata]{\small          
Marginalized one-dimensional distributions for $\Omega_k$ for the baseline
model which allows non-zero curvature.  Weak priors imposed are those outlined in          
Table~\ref{tab:weakpriors}.  We restrict $\Omega_k$ to the range -0.3
and 0.3.  The relatively wide scope for positive curvature is associated
with the angular-diameter-distance degeneracy which is only partly broken
by the CMB data.  Application of the HST prior on $H_0$ to the larger
combined data set somewhat reduces the possibility of significant curvature.              
                \label{fig:curve}}

\end{minipage}
\hfill
\end{figure*}

\cite{Spergel:2003cb} report a detection of the running index of
${\nrun} = -0.031^{+0.016}_{-0.017}$ from their combined WMAPext+2dFGRS+Lyman
$\alpha$ data set.  ~\citet{Slosar:2004fr} present a reduction in
significance of the detection of ${\nrun}$ when their full
likelihood analysis and detailed foreground removal is applied to the
WMAP data.  We find that the Slosar-Seljak modification to WMAP decreases the
significance of ${\nrun}$, but that inclusion of the data from the small scale CMB
experiments has the opposite effect (as was the case found by
\cite{Spergel:2003cb}).  From CMB data alone we determine a median value
for ${\nrun} = -0.072 \pm 0.036$.  This result is somewhat
sensitive to our choice of prior.  \cite{Spergel:2003cb} apply a strong
$\tau < 0.3$ prior which effectively reduces the median value of ${\nrun}$ for their CMB
data only case.  Here we apply a Gaussian HST prior on $H_0$ which lowers the    
significance of the running index to  ${\nrun} =              
-0.065^{+0.035}_{-0.034}$ for the CMBall + \BOOMERANG\ data set.
Inclusion of the LSS data (with uniform prior in $b_g^2$) further 
reduces the significance and our median value from the larger
combined data set is  ${\nrun} = -0.051^{+0.027}_{-0.026} $.  We note that the 
application of a Gaussian prior to $b_g^2$ has no impact on the running
index parameter.  Application of the
HST prior on $H_0$ yields a final median value ${\nrun} =
-0.048 \pm 0.026 $ for the CMBall + \BOOMERANG\ + LSS (+HST) data set.
Figure~\ref{fig:running}
shows the likelihood curves for the ${\nrun}$ parameter for various data
combinations.  It is interesting to compare our result with that of \cite{Seljak:2004xh}, who argue that if the state-of-the-art
modeling of Lyman $\alpha$ forest measurements is dominated by statistical rather than systematic errors,
then $| n_{\rm run} | < 0.01$

\subsubsection{Curvature}
\label{curve}

We consider a modification to the standard model which allows the
possibility of 
non-flat geometry.  We parameterize the curvature
density by $\Omega_k$ and allow it to vary between -0.3 to 0.3.  Table~\ref{tab:mod} shows the results for the CMBall+B03 and CMBall+B03+LSS
data sets.  The CMB data alone places a constraint on the curvature which is
 $\Omega_k = -0.037^{+0.033}_{-0.039} $.
We show in  Figure~\ref{fig:curve} the
likelihood profiles for WMAP, WMAP+B03, CMBall+B03 and CMBall+B03+LSS.
While the addition of \BOOMERANG\ data to the WMAP data tends to lower
the significance of curvature, adding more small scale CMB data
increases the width of the low end tail.  Addition of the LSS data, with
uniform prior in $b_g^2$, yields a median value of  $\Omega_k =
  -0.027 \pm 0.016$.  Application of the Gaussian prior in
$b_g^2$ (with 10\% uncertainty in $b_g$) has a slight effect with a
resulting median value of $\Omega_k = -0.022 \pm 0.015 $. 
If we restrict the $H_0$ value by the
application of a Gaussian HST prior, the curvature density determined from the
CMBall + \BOOMERANG\ data set is  $\Omega_k =
-0.015 \pm 0.016$.  Moreover, application of the more
stringent $H_0$ prior reduces the median value of the curvature from the
combined CMBall + \BOOMERANG\ + LSS data set (flat $b_g^2$ prior) to $\Omega_k =
-0.025 \pm 0.017$.  Our result agrees well with the constraint $\Omega_k =-0.010\pm 0.009$ 
obtained by combining CMB data with the red luminous galaxy clustering
data, which has its own signature of baryon acoustic oscillations \citep{Eisenstein:2005su}. 

\subsubsection{Tensor Modes}
So far we have assumed only scalar perturbations.  However
inflationary models can produce tensor perturbations from
gravitational waves that are predicted to evolve independently of the
scalar perturbations, with an uncorrelated power spectrum ${\cal P}_{\rm
t}$.  The amplitude of a tensor mode falls off rapidly after horizon
crossing and the effect is therefore predominantly on the largest
scales: tensor modes entering the horizon along the line of site to
last scattering distort the photon propagation and generate an
additional anisotropy pattern.  We parameterize the tensor component
by the ratio $A_{\rm t}/A_{\rm s}$, where $A_{\rm t}$ is the
primordial power in the transverse traceless part of the metric tensor
on $0.05\Mpc^{-1}$ scales. We impose a very weak prior on the
amplitude ratio, restricting it to lie between 0 to 20.

A tensor spectral index, defined by ${\cal P}_{\rm t} \propto
k^{n_{\rm t}}$, must also be set. In inflation models it is related
to the amplitude ratio by $A_{\rm t}/A_{\rm s} \approx -8 n_{\rm
t}/(1-n_{\rm t}/2)$, so one parameter suffices. In a nearly uniformly
accelerating regime $n_{\rm t} \approx n_{\rm s} -1$ is also
expected.  However, although $n_{\rm s} >1$ can arise in inflation
models, $n_{\rm t} > 0$ is difficult to obtain.  As well, for many
inflation models with $n_{\rm s}-1$ just below zero, $n_{\rm t}$ comes
out to be slightly closer to zero. If the acceleration changes
significantly over the observably range, $n_{\rm t}$ and $n_{\rm s}$ would not be
intimately tied.  Rather than let $n_t$ float as a second added
parameter, we have chosen to make ${\cal P}_{\rm t}$ flat in $k$ (and
thus set $n_{\rm t}$ to zero) for the computations of the tensor-induced
component of ${\cal C}_\ell$.

\begin{figure*}                                                      
\begin{minipage}[t]{9cm}                                             
\includegraphics[width=0.9\textwidth]{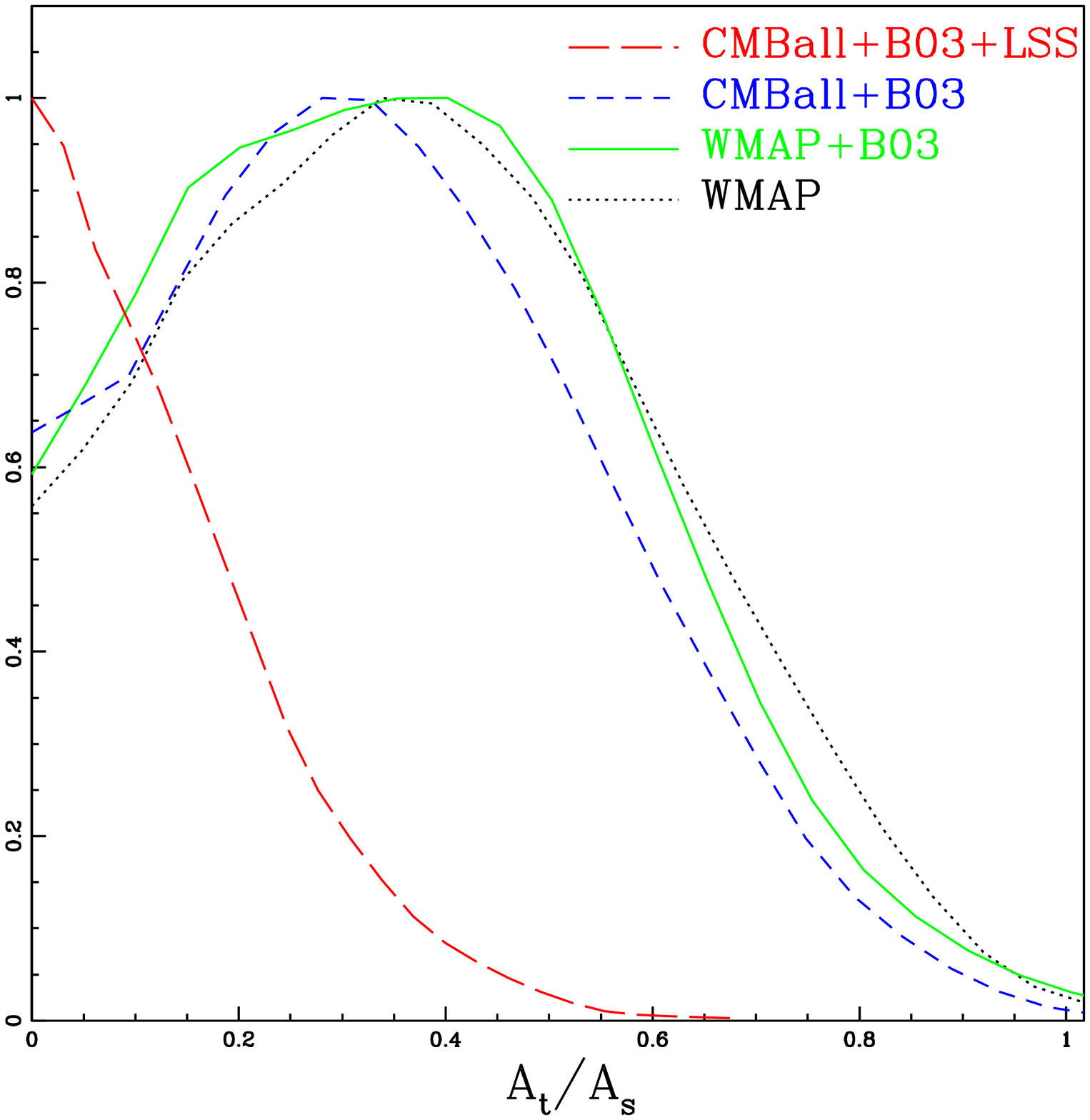}      
\caption[b2kdata]{\small                                             
Marginalized one-dimensional distributions for the amplitude ratio $A_{\rm
t}/A_{\rm s}$ for the baseline                        
model modification which allows tensor modes.  Weak priors imposed are those outlined in                 
Table~\ref{tab:weakpriors}.  We impose the weak prior $0 < A_{\rm
t}/A_{\rm s} < 20$
to the tensor contribution.  We find from CMB data alone                                                    
(CMBall+B03) an upper limit (95\% confidence) on the amplitude ratio of
$A_{\rm t}/A_{\rm s}<0.71$.  For these models however, $H_0$ is only poorly
constrained (see text).  Addition of the LSS data reduces this limit
to  $A_{\rm t}/A_{\rm s}<0.36$.                                       
                        \label{fig:tens}}                         
                                                                     
\end{minipage}                                                       
\hfill                                                               
\begin{minipage}[t]{9cm}                                             
\includegraphics[width=0.9\textwidth]{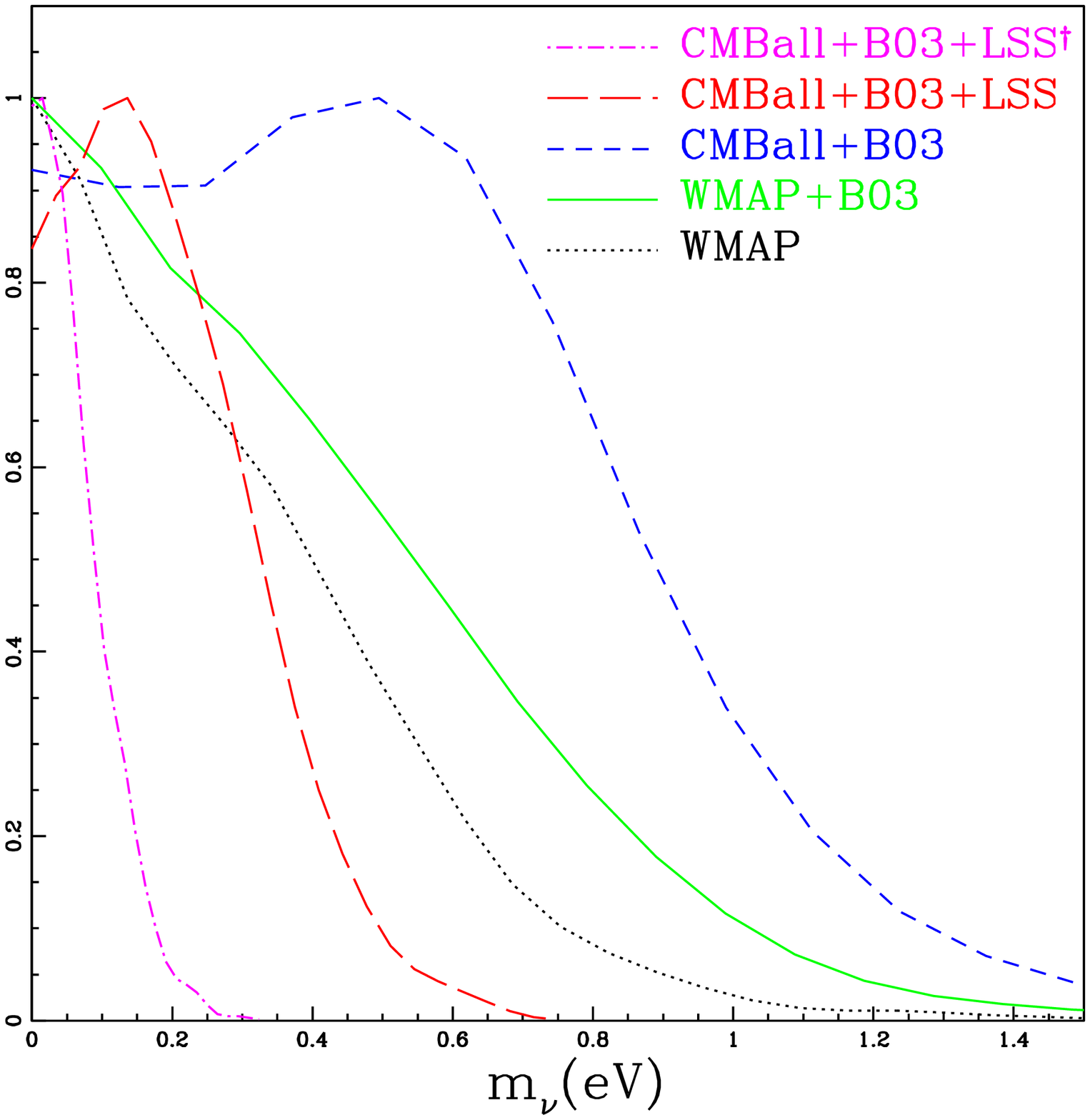}        
\caption[b2kdata]{\small                                             
Marginalized one-dimensional distributions for $m_{\nu}$ for the baseline          
model which allows massive neutrinos (3 species of degenerate mass).  Weak priors imposed are those outlined in 
Table~\ref{tab:weakpriors}.  We parameterize the massive neutrino
contribution as a fraction of the dark matter energy density, $f_{\nu} =
\Omega_{\nu}h^2/\Omega_{DM}h^2$.  We find from CMB data alone
(CMBall+B03) an upper limit (95\% confidence) on the neutrino mass of $m_{\nu} < 1.0
\eV$.  Adding the LSS data reduces this limit to $m_{\nu} <  0.40
\eV$, without any $b_g$ constraint, and to $m_{\nu} < 0.16 \eV$, when $b_g =
1.0 \pm 0.10$ is used. 

                \label{fig:neu}}                                   
                                                                     
\end{minipage}                                                       
\hfill 
\end{figure*}

Results are presented
in Table~\ref{tab:mod}, and Figure~\ref{fig:tens} illustrates the
likelihood curves for the amplitude ratio for a number of data
combinations.  The influence of the high precision of the WMAP data on
the largest scales is evident.  Adding the small scale CMB data only slightly
reduces the limit.  We determine an upper limit on the tensor ratio from
CMB data (CMBall+B03 data set) alone of $A_{\rm t}/A_{\rm s}<0.71$ (95\% confidence limit).
The CMB data appear to select models with relatively large 
tensor-to-scalar ratios.  However, these models which have large values
for the Hubble parameter ($H_0 \sim 85$) are allowed due to the poor
constraint on $H_0$ when including tensor modes.  In this case, the
constraints on $H_0$ are driven mainly by our choice of weak priors and
the data only provides a lower limit (see Table~\ref{tab:mod}).
We find with application of the HST prior (which excludes these models
with large $H_0$ values) that the tensor limit
from the CMBall+B03 data set is reduced to $A_{\rm t}/A_{\rm s}<0.635$.
A similar effect is obtained with the addition of the LSS data which 
further reduces the limit to $A_{\rm t}/A_{\rm s}<0.36$.  When we constrain $b_g$ in the LSS data, the
limits are very similar.  The application of the more restrictive
prior discussed above, with only $n_{\rm s} \leq 1$ allowed to have a
tensor contribution, lowers the CMBall+B03 limit to $A_{\rm t}/A_{\rm
s}<0.45$ and the CMBall+B03+LSS limit to $A_{\rm t}/A_{\rm s}<0.31$.  As
a final case we set $n_{\rm t} = -(A_{\rm t}/A_{\rm s})/8$ and found
$A_{\rm t}/A_{\rm s}<0.54$ for the CMBall+B03
data set and $A_{\rm t}/A_{\rm s}<0.30$ for the combined CMBall+B03+LSS
results (again applying the $b_g$ constraint has little effect).

\subsubsection{Massive Neutrinos}

Observational evidence from solar and atmospheric neutrino experiments,
such as the Sudbury Neutrino Observatory \citep{Ahmad:2002jz} and
Super-Kamiokande \citep{Toshito:2001dk}, suggest that neutrinos change
flavour:  neutrinos of different generations
oscillate into each other.  The implication of flavour changing is that
neutrinos have mass.  Given that neutrinos are the second most abundant
particles in the Universe, massive neutrinos could have considerable
impact on the energy density of the early Universe.  We consider here
the case of three neutrinos of degenerate mass, such that $\Omega_{\nu}h^2
= 3 m_{\nu}/94.0\eV$.  This assumption is well justified given the small
square mass difference measured by oscillation experiments (at most $\delta
m_{\nu}^2 \sim 10^{-3} \eV$,  \cite{Aliani:2003ns}).
We parameterize the massive neutrino contribution as
a fraction of the dark matter energy density, $f_{\nu} =
\Omega_{\nu}h^2/\Omega_{DM}h^2 = 1 - \Omega_{CDM}h^2/\Omega_{DM}h^2$.  

Results for the combined data sets are
given in Table~\ref{tab:mod}.  From CMB data alone the upper limit on
the neutrino fraction is $f_{\nu} < 0.21$ (95\% confidence limit).  This translates to an upper limit on
the neutrino mass of $m_{\nu} < 1.0 \eV$ or $\Omega_{\nu}h^2 < 0.033$.
This limit is more stringent than the $3 \eV$ upper limit on
the electron neutrino mass determined from tritium beta decay
experiments and recommended in the Review of Particle Physics \citep{PDBook}.
Including the LSS data (flat $b_g^2$ prior) pushes this limit down considerably to $f_{\nu} <0.093$
(95\% confidence) and limits the neutrino mass to $m_{\nu}
< 0.40 \eV$.  This result is somewhat larger than that found in
\cite{Spergel:2003cb}.  We find that addition of more and more small
scale CMB data drives the limit up as is evident in
Figure~\ref{fig:neu}.  When $b_g = 1.0 \pm 0.10$ is used, the
neutrino fraction upper limit is reduced to $f_{\nu} <0.041$ (95\% confidence),
corresponding to a neutrino mass limit of $m_{\nu} < 0.16 \eV$. 
This neutrino mass limit is in good
agreement with the strong limit ($m_{\nu} < 0.18 \eV$) obtained by \citep{Seljak:2004sj}, who
included the bias constraint and the SDSS and WMAP data.  In their
analysis of $b_g$ they found $\sigma_8 =0.85^{+0.07}_{-0.06}$, with $b_g
= 1.02^{+0.08}_{-0.08}$.  This
compares with the values we obtain:  $\sigma_8 = 0.85 \pm 0.04$ with
$\delta b_g = 0.10$ and $\sigma_8 = 0.74 \pm 0.08$ with $\delta b_g = \infty$.

\subsubsection{Dark Energy}
\label{darke}

The standard model predicts (and CMB observations strongly support) a
universe which is nearly flat, implying a total energy density
approaching critical.  The total matter density however, comprises
only one third of the total energy density.  The prevailing energy
density component comes from some form of dark energy which up to now
we have assumed takes the form of a vacuum density or cosmological
constant, $\Lambda$, with equation of state described by $w = p/\rho =
-1$, where $p$ and $\rho$ are the dark energy pressure and density
respectively.  We now consider the possibility that the dark energy
component is a rolling scalar field or quintessence (see for example 
 \cite{Ratra:1987rm} or \cite{Huey99}), allowing the effective constant
equation of state parameter $w$ to differ from $-1$. We treat $w$ as a
redshift-independent phenomenological factor and allow it to range
with a uniform prior over the range $-4$ to $0$.  We have also run the
cases with $w$ restricted to lie in the range $-1$ to $0$ and find
similar limits. To be self-consistent, perturbations in the dark
energy should be allowed for when $w$ is not $-1$, although these have
a small impact and only at low multipoles. We set the effective sound speed for
the perturbations to unity in \CAMB, the value for a scalar
field.

\begin{figure*}                                                                                      
\begin{minipage}[t]{9cm}                                                                             
\includegraphics[width=0.9\textwidth]{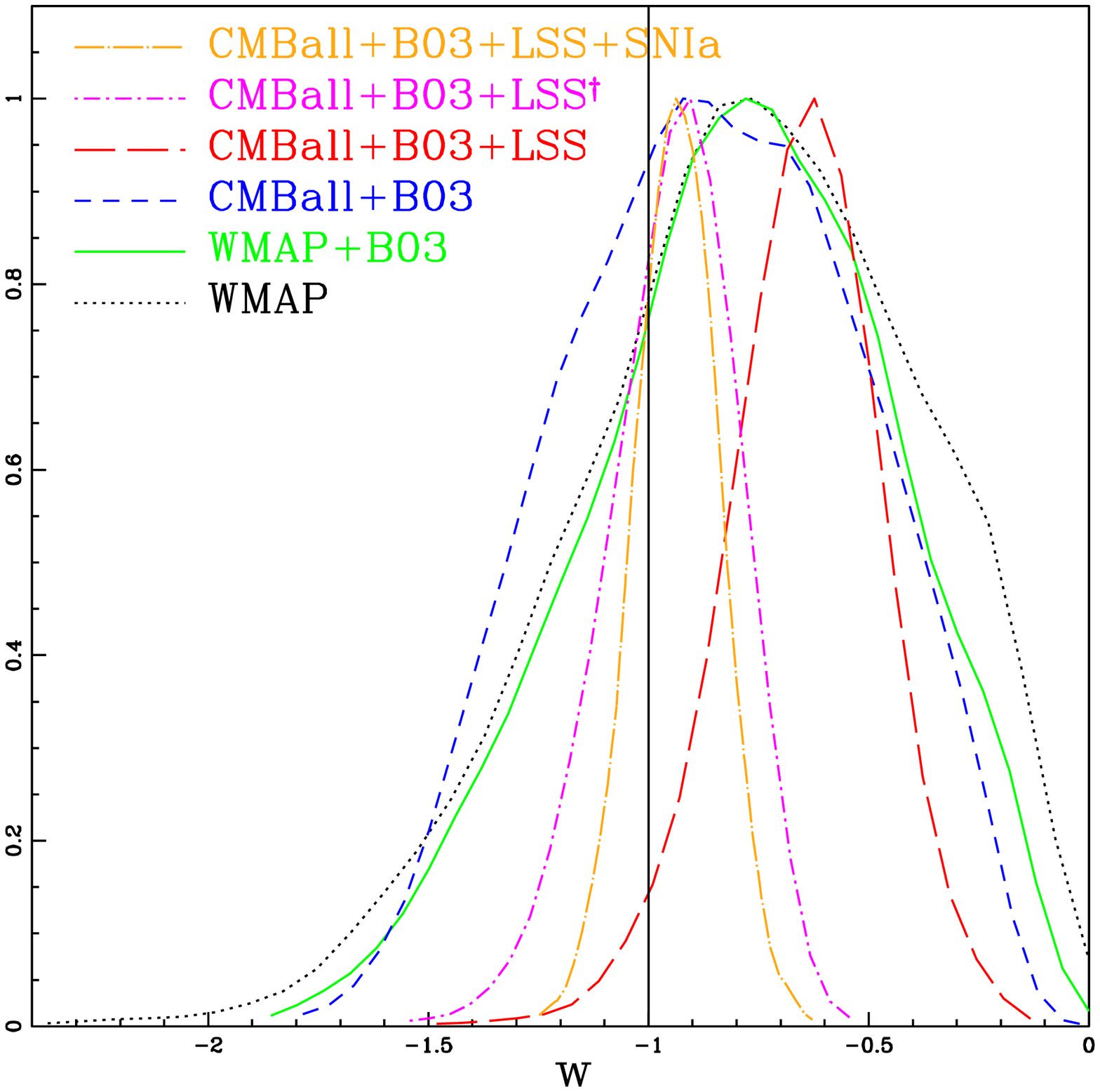}                                     
\caption[w1dplot]{\small Marginalized one-dimensional distributions
for the dark matter equation of state parameter $w$.  Weak priors
imposed on the base parameters are those outlined in
Table~\ref{tab:weakpriors}.  We also impose the prior $-4 < w < 0$.
The $\dagger$ denotes the application of a Gaussian prior to $b_g^2$
(with $b_g = 1 \pm 10\%$).  The nominal flat uniform prior on $b_g^2$
yields a slightly higher median value for $w$, driven by higher values
of $b_g$. Adding the SNIa data however, reduced the median value to $
-0.94 \pm 0.1$.
                        \label{fig:w}}                                                            
                                                                                                     
\end{minipage}                                                                                       
\hfill                                                                                               
\begin{minipage}[t]{10cm}                                                                             
\includegraphics[width=0.9\textwidth]{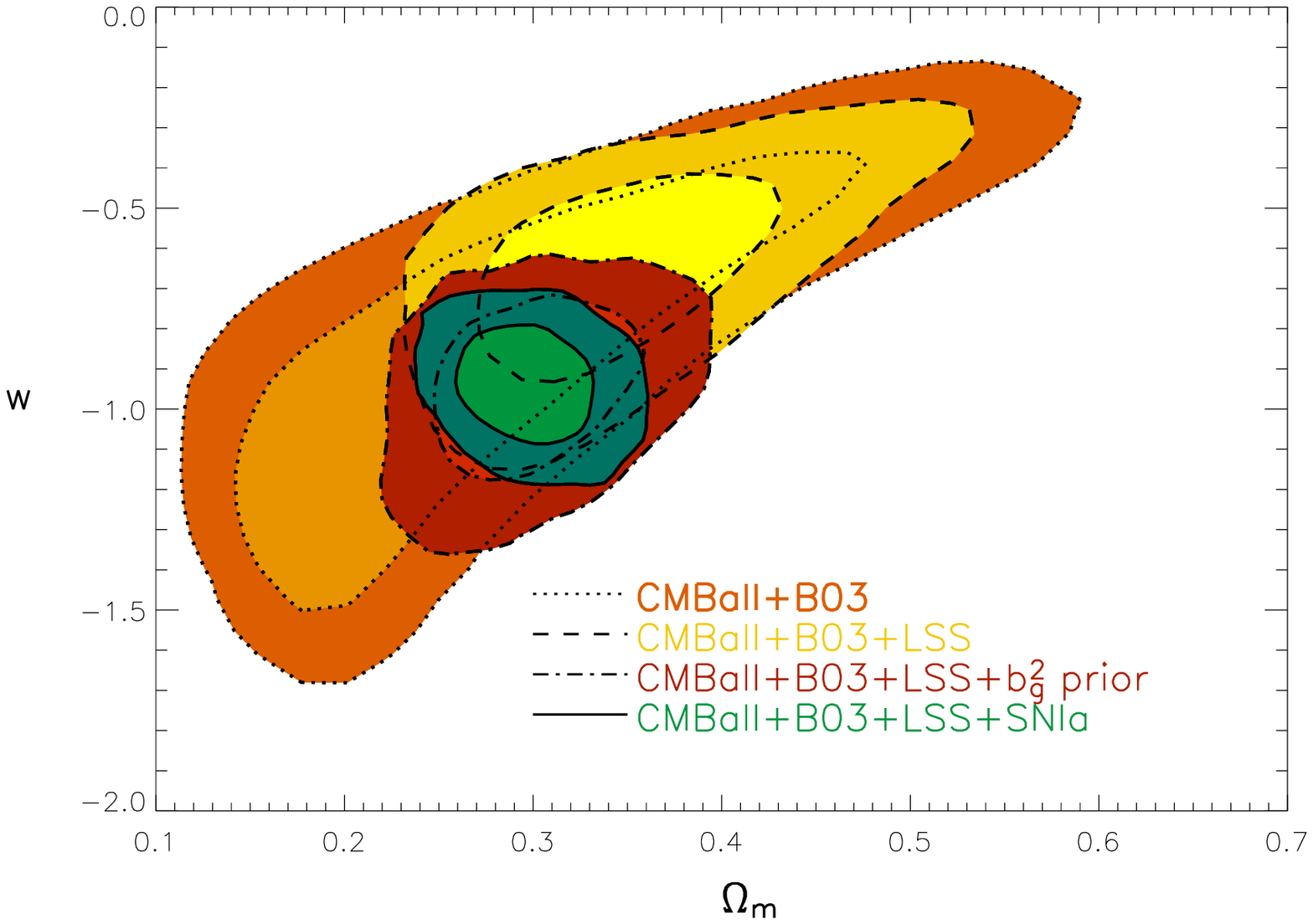}  
\hfill
\caption[w2dplot]{\small 
Constraints on $w$ versus                                                                  
 $\Omega_m$ for a flat $\Lambda$CDM model that allows the dark energy
equation of state parameter, $w$, to differ from -1.  Inner contours represent 68\% likelihood regions and outer
 contours 95\% likelihood regions.  A more stringent Gaussian
$b_g^2$ prior (with $b_g = 1 \pm 10\%$) or the addition of SNIa data is required to
break the strong geometric degeneracy.

                \label{fig:w2d}}                                                                     
                                                                                                     
\end{minipage}                                                                                       
\hfill                                                                                               
 \end{figure*}

The marginalized one-dimensional distributions for
various data combinations are presented in Figure~\ref{fig:w}.  
We find from CMB data alone $w =
-0.86^{+0.36}_{-0.35}$.  The addition of
the LSS data, applying the conservative uniform flat prior on $b_g^2$ to the galaxy
bias factor, yields a median value of $w = -0.65 \pm 0.16$.  
This result is highly sensitive to our choice of prior on $b_g^2$.  The uniform flat
prior on $b_g^2$ gives a relatively high best fit bias value of
$b_g = 1.3$.  Applying a more restrictive Gaussian prior to $b_g^2$ gives:  
$w= -0.95 \pm 0.15 $ with $b_g = 1.0 \pm 0.10$;         
and $w= -0.75 \pm 0.15 $ with $b_g = 1.0 \pm 0.50$.
We explore the effect of adding the SNIa data which significantly                             
improves the constraint on $w$ yielding  $w =  -0.94^{+0.094}_{-0.096} $ with the
flat prior on $b_g^2$.  Results for the CMBall+B03 data set and the CMBall+B03+LSS+SNIa data set
are given in Table~\ref{tab:mod}.  

Figure~\ref{fig:w2d} illustrates the degeneracy in the
$\Omega_m - w$ plane that cannot be broken by CMB data alone and is only
weakly broken with the addition of the LSS data (flat prior on $b_g^2$).
Application of a more restrictive Gaussian prior to $b_g^2$ for the LSS data or addition of
the SNIa data breaks the degeneracy.

\begin{table*}
\centering
\space
\caption{\rm
Marginalized parameter constraints for five modifications of the
baseline model.  Parameter uncertainties represent the 68\% confidence
interval.  For the case of upper or
lower bounds 95\% confidence limits are                                  
quoted.  The following flat weak priors are imposed on the base 6
parameters (as outlined in                                               
Table~\ref{tab:weakpriors}):                                                                             
$0.5 \leq n_{s} \leq 1.5$; $2.7 \leq \ln(10^{10} A_{s})                                                  
\leq 4.0$; $0.005 \leq \Omega_{b}h^{2} \leq 0.1$; $0.01 \leq                                             
\Omega_{c}h^{2} \leq 0.99$; $0.5 \leq \theta \leq 10.0$; and                                             
$0.01 \leq \tau \leq 0.8$.  Additional weak priors restrict                                              
the age of the universe to $10 \Gyr \leq \text{age} \leq 20 \Gyr$ and the                                
expansion rate to $45 \leq H_0 \leq 90$.  We add, in turn, one parameter
to the base set and impose the following prior on each:  running index, $-0.3 < {\nrun} < 0.3$;
curvature, $-0.3 < \Omega_k < 0.3$ ; amplitude ratio, $0 < A_{\rm
t}/A_{\rm s} <
20$; neutrino fraction of dark matter,  $0 < f_{\nu} < 1$; and dark
energy equation of state, $-4 < w < 0$. 
The CMBall data set is as                                      
given in Table~\ref{tab:cmball}.  The LSS data consists of the matter                                    
power spectra from the
2dFGRS and SDSS redshift surveys. We have marginalized the galaxy bias
factor $b_g$ assuming a uniform flat prior in $b_g^2$.  For the $Base+w$
case only we add in both the LSS and SNIa data which gives a better
constrained result than that which is obtained from adding LSS data
alone.  We include the baseline model results
for comparison.
\label{tab:mod}
}
\scriptsize
\begin{tabular}{|c||c|c|c|c|}
\hline\hline
\space
& \multicolumn{2}{c|}{\em Base+Running Index} & \multicolumn{2}{c|}{\em Base+Curvature} \\
 \cline{2-5}
& & & & \\
& CMBall &    CMBall     & CMBall & CMBall\\
& +B03   & +LSS+B03 & +B03   & +LSS+B03 \\
& & & & \\
\hline
& & & & \\
$        \Omega_b h^2$ & $ 0.0238^{+0.0019}_{-0.0019} $ &$ 0.0219^{+0.0010}_{-0.0010} $ & $ 0.0231^{+0.0010}_{-0.0013} $  &$ 0.0227^{+0.0009}_{-0.0010} $   		\\
$        \Omega_c h^2$ & $ 0.104^{+0.015}_{-0.013} $    &$ 0.125^{+0.007}_{-0.007} $    & $ 0.106^{+0.011}_{-0.011} $     &$ 0.111^{+0.008}_{-0.008} $     \\
$          \theta    $ & $ 1.048^{+0.005}_{-0.005} $    &$ 1.044^{+0.004}_{-0.004} $    & $ 1.045^{+0.005}_{-0.005} $     &$ 1.044^{+0.004}_{-0.004} $       	\\
$            \tau    $ & $ 0.31^{+0.13}_{-0.14} $       &$ 0.151^{+0.025}_{-0.036} $    & $ 0.178^{+0.087}_{-0.092} $     &$ 0.139^{+0.021}_{-0.129} $       	\\
$             n_{\rm s}    $ & $ 0.95^{+0.06}_{-0.06} $       &$ 0.90^{+0.04}_{-0.04} $       & $ 0.97^{+0.03}_{-0.04} $        &$ 0.96^{+0.02}_{-0.03} $         	\\
$    \ln[10^{10} A_{\rm s}]$ & $  3.5^{+ 0.2}_{- 0.3} $       &$  3.2^{+ 0.1}_{- 0.1} $       & $  3.2^{+ 0.2}_{- 0.2} $        &$  3.1^{+ 0.1}_{- 0.1} $         	\\
$  \Omega_\Lambda    $ & $ 0.78^{+0.07}_{-0.07} $       &$ 0.67^{+0.04}_{-0.04} $       & $ 0.67^{+0.12}_{-0.13} $        &$ 0.67^{+0.05}_{-0.04} $            \\
$       {\rm  Age(Gyr)}   $ & $ 13.3^{+ 0.4}_{- 0.4} $       &$ 13.7^{+ 0.2}_{- 0.2} $       & $ 14.9^{+ 1.3}_{- 1.3} $        &$ 14.8^{+ 0.7}_{- 0.7} $            \\
$        \Omega_m    $ & $ 0.22^{+0.07}_{-0.07} $       &$ 0.33^{+0.04}_{-0.04} $       & $ 0.37^{+0.17}_{-0.16} $        &$ 0.36^{+0.05}_{-0.06} $            \\
$        \sigma_8    $ & $ 0.91^{+0.07}_{-0.07} $       &$ 0.89^{+0.06}_{-0.06} $       & $ 0.81^{+0.06}_{-0.06} $        &$ 0.82^{+0.06}_{-0.06} $            \\
$          z_{re}    $ & $ 24.7^{+ 5.7}_{- 6.4} $       &$ 16.8^{+ 5.0}_{- 5.0} $       & $ 16.6^{+ 5.7}_{- 5.5} $        &$ 14.7^{+ 4.7}_{- 4.7} $            \\
$            H_0     $ & $ 78.1^{+11.9}_{- 8.7} $       &$ 66.8^{+ 1.3}_{- 1.5} $       & $ 62.5^{+13.1}_{-17.5} $        &$ 61.7^{+ 2.4}_{- 2.1} $            \\
$         {\nrun}    $ & $ -0.072^{+0.036}_{-0.036} $ &$ -0.051^{+0.027}_{-0.026} $  &	-		     &                 -              	\\
$        \Omega_k    $   &	-			  &	-			  & $ -0.037^{+0.033}_{-0.039} $  & $ -0.027^{+0.016}_{-0.016}$  \\	
& & & & \\
\hline\hline
\space
& \multicolumn{2}{c|}{\em Base+Tensor Modes} & \multicolumn{2}{c|}{\em
Base+Massive Neutrinos} \\
 \cline{2-5}
& & & & \\
& CMBall &    CMBall     & CMBall & CMBall\\
& +B03   & +LSS+B03 & +B03   & +LSS+B03 \\
& & & & \\
\hline
& & & & \\
$        \Omega_b h^2$ &  $ 0.0246^{+0.0013}_{-0.0013} $ & $ 0.0232^{+0.0009}_{-0.0009} $    & $ 0.0226^{+0.0014}_{-0.0014} $&  $ 0.0224^{+0.0008}_{-0.0009} $  \\
$        \Omega_c h^2$ &  $ 0.0957^{+0.0090}_{-0.0086} $ & $ 0.117^{+0.006}_{-0.006} $       & $ 0.120^{+0.014}_{-0.014} $   &  $ 0.126^{+0.007}_{-0.007} $      \\
$          \theta    $ &  $ 1.048^{+0.004}_{-0.004} $    & $ 1.046^{+0.004}_{-0.004} $       & $ 1.047^{+0.005}_{-0.005} $   &  $ 1.045^{+0.004}_{-0.004} $      \\
$            \tau    $ &  $ 0.162^{+0.072}_{-0.070} $    & $ 0.104^{+0.049}_{-0.049} $       & $ 0.164^{+0.094}_{-0.088} $   &  $ 0.108^{+0.049}_{-0.047} $     \\
$       n_{\rm s}    $ &  $ 1.02^{+0.04}_{-0.04} $       & $ 0.97^{+0.02}_{-0.02} $          & $ 0.95^{+0.04}_{-0.04} $      &  $ 0.95^{+0.02}_{-0.02} $         \\
$\ln[10^{10} A_{\rm s}]$& $  3.1^{+ 0.1}_{- 0.1} $       & $  3.1^{+ 0.1}_{- 0.1} $          & $  3.2^{+ 0.2}_{- 0.2} $      &  $  3.1^{+ 0.1}_{- 0.1} $         \\
$  \Omega_\Lambda    $ &  $ 0.82^{+0.03}_{-0.04} $       & $ 0.72^{+0.03}_{-0.03} $          & $ 0.64^{+0.11}_{-0.11} $      &  $ 0.64^{+0.06}_{-0.05} $         \\
$     {\rm Age(Gyr)} $ &  $ 13.2^{+ 0.3}_{- 0.2} $       & $ 13.5^{+ 0.2}_{- 0.2} $          & $ 14.1^{+ 0.4}_{- 0.4} $      &  $ 13.9^{+ 0.2}_{- 0.2} $         \\
$        \Omega_m    $ &  $ 0.181^{+0.036}_{-0.034} $    & $ 0.28^{+0.03}_{-0.03} $          & $ 0.36^{+0.11}_{-0.11} $      &  $ 0.36^{+0.05}_{-0.06} $         \\
$        \sigma_8    $ &  $ 0.77^{+0.07}_{-0.07} $       & $ 0.84^{+0.05}_{-0.05} $          & $ 0.60^{+0.13}_{-0.12} $      &  $ 0.74^{+0.08}_{-0.08} $         \\
$          z_{re}    $ &  $ 15.2^{+ 4.5}_{- 4.5} $       & $ 12.2^{+ 4.0}_{- 4.0} $          & $ 16.6^{+ 6.2}_{- 5.9} $      &  $ 13.0^{+ 4.1}_{- 4.0} $         \\
$            H_0     $ &  $ > 73.2                 $     & $ 71.4^{+ 2.8}_{- 2.9} $          & $ 64.9^{+ 8.1}_{- 7.4} $      &  $ 64.8^{+ 3.9}_{- 3.8} $         \\
$ A_{\rm t}/A_{\rm s}$ &  $ <0.71                  $     & $ <0.36                $         & -			&	-			\\
$        f_{\nu}    $  & -			& -				&  $ <0.21$  & $ <0.09 $  \\
& & & & \\
\hline\hline                                                                                                                                                        
\space
& \multicolumn{2}{c|}{\em Base+w} & \multicolumn{2}{c|}{\em Baseline} \\                                                                                                                                          
 \cline{2-5}    
& & & & \\                                                                                                                                                    
& CMBall &    CMBall     & CMBall & CMBall\\                                                                                                                        
& +B03   & +LSS+B03+SNIa & +B03   & +LSS+B03 \\                                                                                                                
& & & & \\
\hline    
& & & & \\                                                                                                                                                          
$        \Omega_b h^2$ & $ 0.0234^{+0.0013}_{-0.0013} $  &$ 0.0229^{+0.0009}_{-0.0009} $   & $ 0.0233^{+0.0013}_{-0.0012} $   &  $ 0.0227^{+0.0008}_{-0.0008} $    \\    
$        \Omega_c h^2$ & $ 0.106^{+0.011}_{-0.011} $     &$ 0.117^{+0.007}_{-0.008} $      & $ 0.106^{+0.010}_{-0.010} $      &  $ 0.120^{+0.005}_{-0.005} $        \\   
$          \theta    $ & $ 1.046^{+0.005}_{-0.005} $     &$ 1.045^{+0.004}_{-0.004} $      & $ 1.045^{+0.004}_{-0.004} $   	 &  $ 1.045^{+0.004}_{-0.004} $        \\   
$            \tau    $ & $ 0.161^{+0.078}_{-0.077} $     &$ 0.118^{+0.053}_{-0.055} $      & $ 0.170^{+0.090}_{-0.081} $    	 &  $ 0.106^{+0.047}_{-0.048} $       \\    
$       n_{\rm s}    $ & $ 0.98^{+0.04}_{-0.04} $        &$ 0.96^{+0.02}_{-0.02} $         & $ 0.98^{+0.04}_{-0.03} $         &  $ 0.95^{+0.02}_{-0.02} $           \\   
$\ln[10^{10} A_{\rm s}]$&$  3.2^{+ 0.1}_{- 0.1} $        &$  3.1^{+ 0.1}_{- 0.1} $         & $  3.2^{+ 0.2}_{- 0.2} $       	 &  $  3.1^{+ 0.1}_{- 0.1} $           \\   
$  \Omega_\Lambda    $ & $ 0.71^{+0.10}_{-0.11} $        &$ 0.70^{+0.02}_{-0.02} $         & $ 0.77^{+0.05}_{-0.05} $       	 &  $ 0.70^{+0.03}_{-0.03} $           \\   
$        {\rm Age(Gyr)} $ & $ 13.7^{+ 0.4}_{- 0.4} $        &$ 13.6^{+ 0.2}_{- 0.2} $         & $ 13.5^{+ 0.2}_{- 0.3} $       	 &  $ 13.6^{+ 0.2}_{- 0.2} $           \\   
$        \Omega_m    $ & $ 0.29^{+0.11}_{-0.10} $        &$ 0.30^{+0.02}_{-0.02} $         & $ 0.23^{+0.05}_{-0.05} $       	 &  $ 0.30^{+0.03}_{-0.03} $           \\   
$        \sigma_8    $ & $ 0.76^{+0.15}_{-0.15} $        &$ 0.82^{+0.06}_{-0.06} $         & $ 0.83^{+0.06}_{-0.06} $       	 &  $ 0.85^{+0.05}_{-0.05} $           \\   
$          z_{re}    $ & $ 15.9^{+ 5.0}_{- 4.9} $        &$ 13.3^{+ 4.2}_{- 4.2} $         & $ 16.5^{+ 5.4}_{- 5.1} $       	 &  $ 12.6^{+ 3.9}_{- 4.0} $           \\   
$            H_0     $ & $ 69.9^{+13.1}_{-12.9} $        &$ 68.6^{+ 2.1}_{- 2.0} $         & $ 75.8^{+ 5.6}_{- 5.1} $      	 &  $ 69.6^{+ 2.4}_{- 2.4} $           \\   
$             w      $ & $ -0.86^{+0.36}_{-0.35} $ &$ -0.94^{+0.094}_{-0.096} $      & -                  &       -                       \\                  
& & & & \\
\hline\hline
\end{tabular}
\centering
\end{table*}

\section{Sub-dominant Isocurvature Model}
\label{iso}

CMB anisotropies provide a powerful probe of the nature of early
universe perturbations. However almost any TT power spectrum shape can
be fit rather well by using contrived combinations of initial
perturbations, for example by adding structure to the primordial power
spectrum, and/or by adding isocurvature modes (which come in many
varieties). With full freedom, determination of the basic cosmic
parameters suffers because of high correlation with these extra
degrees of freedom.  These degeneracies are broken by the addition of
polarization data, because of the pattern differences of the peaks and
troughs between EE, TE and TT power spectra. In particular the
intensity and polarization power spectra for isocurvature modes have
the peaks out of phase with those from adiabatic modes (see {\it
e.g.}, \citet{Bond:1987ub},~\citet{Hu:1996qs}). A mix of isocurvature
and adiabatic modes can be designed to give acceptable fits to the CMB
intensity power spectra ({\it e.g.},
\citep{Bucher:2004an,Kurki-Suonio:2004mn}), and the polarization data,
from B03 as well from DASI, CAPMAP and CBI are
not yet at the point to clearly distinguish among these more complex
models.

To illustrate the constraints that can be determined from the current
CMB data, we consider here a simple hybrid case consisting of our
basic adiabatic mode model with constant spectral index, a single cold
dark matter (CDM) isocurvature mode with its own constant primordial
spectral index $n_{\rm iso}$, with no correlation between the
two. This adds another two parameters to our basic six, $n_{\rm iso}$
and an amplitude ratio $R_2 \equiv (A_{iso}/A_{\rm s})$.  We assume
the isocurvature perturbations are Gaussian-distributed as we have
done for the adiabatic modes.  Results are shown in
Table~\ref{tab:iso} for the CMBall+B03+HST data combination.  Aside
from the more stringent HST data prior on $H_0$, all priors on the 6
base parameters are as outlined in Table~\ref{tab:weakpriors}.
Although results indicate that there is no evidence for the presence
of an isocurvature mode, the upper limits still allow for a
sub-dominant component.

We now expand on the theoretical framework. Isocurvature modes may
arise in two (or more) field models of inflation, and, though
certainly not a natural prediction, they have reasonably good physical
motivation.  Isocurvature modes could also be generated after
inflation ended. An ingredient needed for isocurvature modes to have
an observable impact on the CMB is that they are associated with a
component of significant mass-energy. If the dark matter is cold and
of one type there are two distinct matter isocurvature modes, the
baryon and CDM modes, involving primordial fluctuations in the
entropy-per-baryon or the entropy-per-CDM-particle.  The classic
example of a CDM possibility is the isocurvature axion mode. It turns
out that the isocurvature CDM and baryon modes actually have almost
identical signatures in the CMB~\citep{Gordon:2002gv}, and hence
cannot be constrained separately.  In this paper we
choose to constrain only the CDM isocurvature mode. There are other
possibilities for isocurvature modes that we will not explore
here\footnote{One set involves neutrinos and photons compensating
each other~\citep{Bucher:1999re,Rebhan:1994zw,Lewis:2004kg}, but these must
be produced after neutrino decoupling and so seem quite unlikely. Much
better motivated are isocurvature modes associated with defects such
as cosmic strings created in early universe phase transitions, which
would contribute to a sub-dominant mass-energy content in the present
epoch. Cosmic-defect-induced perturbations are greatly modified from
their largely-uncorrelated initial state through gravity wave emission
and have distinctively non-Gaussian features. By themselves they
cannot explain the CMB data since the peaks and troughs are difficult
to mimic, but they could be a sub-dominant component that future CMB
data should be able to constrain.}.

It is a straightforward modification of \COSMOMC\ to include an
additional isocurvature component in the MCMC chains, which uses \CAMB\
to compute the isocurvature power spectra. Instead of using $n_{\rm iso}$
as a basic parameter, we use two amplitude ratios for the two
parameters that characterize our CDM isocurvature mode, following a
suggestion of \citet{Kurki-Suonio:2004mn}:
\begin{eqnarray}
&& R_2 \equiv {\cal P}_{\rm iso}(k_2)/ {\cal P}_{\rm s} (k_2)\,  , \
k_2=0.05\Mpc^{-1},  \\
&& R_1 \equiv {\cal P}_{\rm iso}(k_1)/ {\cal P}_{\rm s} (k_1), \
k_1=0.005\Mpc^{-1} \, . \nonumber
\end{eqnarray}
Here ${\cal P}_{\rm s}(k)$ is the power in the primordial curvature
perturbation and ${\cal P}_{\rm iso}(k)$ is the power in the
primordial CDM-photon entropy perturbation.  The $k_2$ scale
corresponds to $\ell \sim 700$ and $k_1$ to $\ell \sim 70$.  We adopt
a uniform prior probability over the range 0 to 20 for $R_1$, and over
0 to 100 for $R_2$.  The isocurvature spectral index $n_{\rm iso}$,
defined by ${\cal P}_{\rm iso}(k) \propto k^{n_{\rm iso}}$, is now a
derived parameter, expressible in terms of $R_1$, $R_2$ and the
adiabatic spectral index $n_{\rm s}$, defined by ${\cal P}_{\rm
s}(k) \propto k^{{n_{\rm s}}-1}$,
\begin{eqnarray}
&& n_{\rm iso} = n_{\rm s} -1 + \ln (R_2/R_1)/ \ln (k_2/k_1) \, .
\end{eqnarray}
Two interesting limiting cases are the scale invariant $n_{\rm iso} = 0$
spectrum with high spatial correlation and the $n_{\rm iso} = 3$
``isocurvature seed'' white noise spectrum with no spatial
correlation\footnote{Such a spectrum is so steep that it must be
regulated by a cutoff at high $k >> k_2$. Physically this is typically
the scale of the horizon when they are generated, but it is
constrained by small scale structure information such as the allowed
epoch of first star formation, which cannot be too early or else
$\tau$ could be far too large. We do not add this cutoff to our study
since the CMB has a larger natural damping scale. A traditional seed
case that has been considered is primordial black hole production.}.

The limits shown in Table~~\ref{tab:iso} demonstrate that the large
scale $R_1$, dominated by the WMAP data, is much better constrained at
$<0.3$ than the small scale $R_2 < 2.3$, which \BOOMERANG\
probes. This translates into a preference for steeper $n_{\rm iso}$ than
the scale invariant value. Since for neither is there an indication of
a non-zero value, just upper limits, the results are sensitive to the
prior probabilities we assign them. Our choice of uniform prior for
$R_1$ and $R_2$ is conservative in that the upper limits decrease with
other choices, {\it e.g.}, one uniform in $\ln (R_i)$ (a
non-informative prior), or one uniform in $n_{\rm iso}$ and $R_2$. The
conservative choice actually downgrades the probability of steep
$n_{\rm iso}$. (The B03pol data by itself only limits $R_1 < 17$ and $R_2 <
22$; the full B03 data gives $R_1 < 1.8$ and $R_2 < 5.3$.)

Further constraints on isocurvature modes arise from LSS
since the shape of the isocurvature matter power spectrum differs in
significant ways from the adiabatic one, but we do not consider those
here.

The strongest constraints come from the low $\ell$ part of the spectrum\footnote{This is largely because of the isocurvature effect                                 
\citep{1986MNRAS.218..103E}: to have no overall energy density perturbation on large         
scales, the entropy perturbation is carried almost entirely by the CDM                       
or baryons, but when the equation of state changes from radiation to                         
matter dominance the photons carry the perturbation. The result is a                         
rather dramatic amplification in the power over the Sachs-Wolfe power                        
familiar in the adiabatic case by 36.  This is why the nearly scale                          
invariant case for isocurvature alone has been ruled out for so long                         
from CMB data, so there was only room for it appearing at a                                  
sub-dominant level.}.
 However, spectra that are significantly steeper than
inflation-motivated nearly-scale-invariant ones are still allowed by the
data.  
To focus attention on the role played by the new, high $\ell$ 
\BOOMERANG\ results, we now fix $n_{\rm iso}$
at 3, the white noise `seed' spectrum, the limiting case in which the
isocurvature perturbations when created were uncorrelated
spatially. The large angular scales are highly suppressed and the isocurvature peaks
and troughs emerge looking somewhat like an $\ell$-shifted version of
the adiabatic spectrum. The two spectra then test at what level
interleaved isocurvature peaks are allowed by the CMB data. Results
are shown in Table~\ref{tab:iso}.  To relate the $R_2 < 3.0$ limit
to a more intuitive expression of what the CMBall+B03+HST data set
data allows, we note that over a bandpower in $\ell$ from 75 to 1400,
${\cal C}_B^{TT({\rm iso})} / {\cal C}_B^{TT({\rm s})} \sim (0.005)R_2$,
hence the upper limit corresponds to an allowed CMB contamination of
this sub-dominant component of TT of only a few percent.  Over a
bandpower in $\ell$ from 150 to 1000 we find ${\cal C}_B^{EE(\rm iso)}
/ {\cal C}_B^{EE(\rm s)} \sim (0.008)R_2$, for the allowed EE
isocurvature bandpower contamination.  B03pol gives $R_2 < 58$. The
full B03 data set including TT gives $R_2 < 9.5$\footnote{The
$n_{\rm iso}=2$ case mimics even more the peak/trough patterns in
${\cal C}_\ell$ except for the shift, so we tested that case as well.
CMBall+B03+HST gives $R_2 < 2.7$, B03 alone, but with TT, gives $R_2 <
6.8$ and B03pol gives $R_2 < 41$.  Translation to the allowed
contamination is done with the bandpower ratios ${\cal C}_B^{TT(\rm
iso)} / {\cal C}_B^{TT(\rm s)} \sim (0.007)R_2$, ${\cal C}_B^{EE(\rm iso)}
/ {\cal C}_B^{EE(\rm s)} \sim (0.009)R_2$.}.

Since the isocurvature models adopted for these tests are not
especially well-motivated physically, we have also chosen not to apply
the LSS prior to our results.

The $n_{\rm iso}=3, 2$ illustration allows us to conclude that even
with the errors on the EE and TE data, there is evidence against the
isocurvature shifted pattern over the adiabatic pattern and only
restricted room for an interleaved peak pattern, at a level below
50\%.  This test differs from the adiabatic-only peak/trough pattern
shift using B03pol Fig.\ref{fig:2Dlikecurves} since there are no
interleaved peaks and troughs in that case.  Examination of the \CAMB\
models obtained from the marginalized constraints in
Table~\ref{tab:iso} reveals that the parameters chosen by \COSMOMC\
adjust to make the adiabatic ${\cal C}_{\ell}^{\rm s}$ pattern
compensate for the isocurvature ${\cal C}_{\ell}^{\rm iso}$
contamination.

\begin{table*}
\space
\caption{\rm                                                                                             
Marginalized parameter constraints for a model which includes both
(dominant) adiabatic and (sub-dominant) isocurvature modes.
Parameter uncertainties represent the 68\% confidence
interval.  Upper bounds are 95\% confidence limits.
The flat weak priors are imposed on the base 6
parameters are as outlined in Table~\ref{tab:weakpriors}. 
The CMBall data set is defined in Table~\ref{tab:cmball}.
We include the baseline model result (with the more stringent HST prior)
for comparison.  We consider two parameterizations for the isocurvature
model.  For the first (column two) we add two parameters to our basic
six:  $R_2 \equiv {\cal P}_{\rm iso}(k_2)/ {\cal P}_{\rm s} (k_2)$, with pivot scale $k_2
=0.05\Mpc^{-1}$; and $R_1 \equiv {\cal P}_{\rm iso}(k_1)/ {\cal P}_{\rm s} (k_1)$ with  $k_1 =
0.005\Mpc^{-1}$.  We impose the priors $0 < R_1 < 20$ and $0 < R_2 <
100$.  For this case the isocurvature spectral index,
$n_{\rm iso}$, is a derived parameter.  We also consider the ``white
isocurvature'' case (column three) where we fix $n_{\rm iso} = 3$ and allow the
amplitude ratio $R_2 \equiv (A_{\rm iso}/A_{\rm s})$ to lie anywhere between 0 and
100.
\label{tab:iso}                                                                                          
                                                                                                        }                                             
\scriptsize                                                                                                                                                                                                                                 
\begin{tabular}{|c||c|c|c|}                                                                                                                                                                                                             
\hline\hline   
& Baseline & Adiabatic + Iso & Adiabatic + White Iso   \\    
\hline   
& & & \\                                                                                                                                                                                                                       
& CMBall & CMBall & CMBall              \\                                                                                                                                                                          
& +B03+HST & +B03+HST & +B03+HST  \\                                                                                                                                                                                                                    
& & & \\
\hline  
& & & \\                                                                                                                                                                                                                                    
$        \Omega_b h^2$ &$ 0.0230^{+0.0010}_{-0.0010} $ &$ 0.0247^{+0.0014}_{-0.0014} $ & $ 0.0235^{+0.0012}_{-0.0011} $     \\          
$        \Omega_c h^2$ &$ 0.108^{+0.008}_{-0.008} $    &$ 0.103^{+0.009}_{-0.009} $    & $ 0.107^{+0.009}_{-0.009} $            \\      
$          \theta    $ &$ 1.045^{+0.004}_{-0.004} $    &$ 1.051^{+0.005}_{-0.005} $    & $ 1.046^{+0.005}_{-0.005} $            \\      
$            \tau    $ &$ 0.151^{+0.069}_{-0.067} $    &$ 0.163^{+0.071}_{-0.069} $    & $ 0.156^{+0.071}_{-0.071} $            \\      
$             n_{\rm s}    $ &$ 0.97^{+0.03}_{-0.03} $ &$ 1.01^{+0.04}_{-0.04}  $ & $ 0.97^{+0.03}_{-0.03} $                 \\          
$             R_1    $ &$ -                    $       &$ <0.28                 $      & $    -                 $         \\      
$             R_2    $ &$ -                    $       &$ <2.3                  $      & $ <3.0                 $               \\      
$       n_{\rm iso}  $ &$ -                    $       &$ 1.1^{+ 0.6}_{- 0.5} $    & $ 3.0\space(fixed)     $ 	           \\          
$  \ln[10^{10} A_{\rm s}]  $ &$  3.1^{+ 0.1}_{- 0.1} $       &$   3.2^{+ 0.1}_{- 0.1} $ & $  3.1^{+ 0.1}_{- 0.1} $              \\          
$  \Omega_\Lambda    $ &$ 0.76^{+0.04}_{-0.04} $       &$ 0.80^{+0.04}_{-0.04} $       & $ 0.77^{+0.04}_{-0.04} $           \\          
$        {\rm Age(Gyr)}  $ &$ 13.5^{+ 0.2}_{- 0.2} $       &$ 13.1^{+ 0.3}_{- 0.3} $       & $ 13.4^{+ 0.2}_{- 0.2} $           \\          
$        \Omega_m    $ &$ 0.24^{+0.04}_{-0.04} $       &$ 0.20^{+0.04}_{-0.04} $       & $ 0.23^{+0.04}_{-0.04} $           \\          
$          z_{re}    $ &$ 15.4^{+ 4.7}_{- 4.5} $       &$ 15.5^{+ 4.4}_{- 4.4} $       & $ 15.6^{+ 4.8}_{- 4.6} $           \\          
$           H_0      $ &$ 74.4^{+ 4.0}_{- 3.9} $       &$ 80.3^{+ 5.6}_{- 5.5} $       & $ 75.9^{+ 4.3}_{- 4.2} $           \\          
\hline\hline                                                                                                                                                                                                                                
\end{tabular}                                                                                                                                                                                                                               
\centering                                                                                                                                                                                                                                  
\end{table*}

\section{Conclusions}

The \BOOMERANG\ data set does well at constraining the cosmological
parameters of the standard $\Lambda$CDM model.  The results are in good
agreement with those derived from other CMB experiments, as is evident
in Table~\ref{tab:base}.  The parameter constraints derived from the
\BOOMERANG\ data set in combination with
the WMAP data are highly competitive with those from the CMBall
data set.  

We have applied the Slosar-Seljak modification                                          
to the WMAP data which has the general effect of broadening slightly the         
WMAP 1D likelihood curves with the largest impacts on the (WMAP alone) median values of          
$\tau$ ($\sim 0.3\sigma$ increase) and $\Omega_m$ ($\sim 0.6\sigma$                
decrease) for the baseline model and  ${\nrun}$ ($\sim 0.5\sigma$                 
increase) for the baseline+running index model.  
Our graphical representation of the standard model parameters in                                         
Figure~\ref{fig:mean} illustrates the impact of the addition of the                                      
LSS data, which shifts the median values very little.  
Figure~\ref{fig:mean} shows the best estimates for the parameters of the   
$\Lambda$CDM model from current CMB and LSS redshift survey data. 

Our analysis of five extensions to the standard model is summarized in         
Table~\ref{tab:mod}.  Intriguingly we found two cases, the running index
and neutrino fraction, where the addition
of data from higher multipole CMB experiments drove median values 
away from the conventional value.
The evidence for a running index however, is
slight ($< 2\sigma$). Our neutrino mass limit (assuming three species
of nearly degenerate mass) from CMB data alone is  
$m_{\nu} < 1.0 \eV$ and from the combined CMB and LSS data set is
$m_{\nu} < 0.40 \eV$, with no $b_g$ prior, and $m_{\nu} < 0.16 \eV$, with $b_g = 1.0 \pm 0.10$.

We have explored the sensitivity of our CMBall+B03+LSS results to the
prior imposed on the galaxy bias factor $b_g$ and have found that only
the neutrino mass results (above) and dark energy equation of state results are
significantly impacted.  For the dark energy equation of state we find
that applying a more restrictive Gaussian prior to $b_g^2$ gives       
$w = -0.97 \pm 0.12$ with $b_g = 1.0 \pm 0.10$.  Without any $b_g$
constraint we find  $w = -0.65 \pm 0.16$.  However, addition of the SNIa data, with
the flat prior on $b_g^2$,  yields $w = -0.94^{+0.094}_{-0.096}$. The
results are consistent with the dark energy being the cosmological
constant. 

While the polarization data is not yet at
the level of accuracy of the intensity data, cross checks of best
fit parameters from the \BOOMERANG pol data and \BOOMERANG TT data indicate 
consistent results.  The consistency of the shape parameter $\theta$ determined
from \BOOMERANG pol and from \BOOMERANG TT demonstrates that the 
peak and trough positions forecast by the spectra are in robust
agreement. Isocurvature modes are beginning to be constrained by the current CMB                
polarization data and our upper limits and phenomenological discussion          
represent a good starting point for future analysis of these more complex models. 
The CMB polarization data is emerging but is not yet driving
parameter determination.  We look forward to future higher precision data
in which polarization data will play a larger role.

\acknowledgements

We gratefully acknowledge support from the CIAR, CSA, and NSERC in Canada, ASI,
University La Sapienza and PNRA in Italy, PPARC and the Leverhulme
Trust in the UK, and NASA (awards NAG5-9251 and NAG5-12723) and NSF
(awards OPP-9980654 and OPP-0407592) in the USA. Additional support
for detector development was provided by CIT and JPL. CBN acknowledges
support from a Sloan Foundation Fellowship, WCJ and TEM were partially
supported by NASA GSRP Fellowships. Field, logistical, and flight
support were supplied by USAP and NSBF; data recovery was particularly
appreciated. This research used resources at NERSC, supported
by the DOE under Contract No. DE-AC03-76SF00098, and the MacKenzie
cluster at CITA, funded by the Canada Foundation for Innovation. We
also thank the CASPUR (Rome-ITALY) computational facilities and the
Applied Cluster Computing Technologies Group at the Jet Propulsion
Laboratory for computing time and technical support. Some of the
results in this paper have been derived using the HEALPix
package~[\cite{Gorski:2004by}] as well as the FFTW package~[\cite{fftw}].  Thanks to Anze Slosar for making his low-$\ell$ WMAP likelihood
code available to us.  The Boomerang
field team is also grateful to the Coffee House at McMurdo
Station, Antarctica, for existing.

\acknowledgments

\bibliographystyle{apj}
\bibliography{b2kpars.bib}

\end{document}